%% file: Swissi-2026-09-guex5c23zfm5-WealthManagementAI-NestedDelegation.tex
\documentclass[oneside,10pt,english]{swissi}
\input{1-system/0-meta.tex}

\title{Multi-Agent AI as a Nested Principal-Agent Problem\\in Private Wealth Management \\[0.35em]
\normalsize \papersubtitle}

\author{{\fontsize{8}{10}\selectfont\paperauthors}\\[0.9em]
{\scriptsize\paperaffiliations}}
\date{\vspace{1.5em}}

\usepackage{flafter}
\usepackage[breakable,skins]{tcolorbox}

\AtEveryBibitem{\iffieldequalstr{journaltitle}{Swissi AI Journal}{\clearfield{doi}}{}}

\AtBeginEnvironment{tabularx}{\footnotesize}
\AtBeginEnvironment{tabular*}{\footnotesize}
\AtBeginEnvironment{algorithmic}{\swAlgSmall}

\begin{document}
\maketitle
\begin{abstract}
In private wealth management, a manager delegating to artificial intelligence (AI) acts as the client's agent and the system's principal. We introduce a model-independent formulation that combines nested principal--agent delegation with constrained joint maximisation as the task assigned to the AI system. The objective represents client and manager outcomes separately over portfolio--workflow pairs. Legal duties, mandate requirements and evidence sufficiency determine admissibility, with Switzerland, Germany and Austria supplying the legal context. Weights and reference-service floors make the trade-off explicit; concession accounting separates their effects on the client. Analytical constructions and a simulation using public-market observations illustrate the approach. Across eight decision states from four constructed mandates, omitted client liabilities caused two liquidity violations, omitted manager terms caused two capacity violations, and mistranslated weights changed four otherwise admissible choices under faithful optimisation. At the declared weights, six states selected a higher service tier than the client-best alternative, with client concessions of EUR~1,178 to EUR~2,264 and manager gains of EUR~3,062 to EUR~10,381. Three instruction forms each reached all 32 specified decisions under shared numerical, evidence and simulated approval controls; professional instructions matched explicit nested delegation on accuracy and clarification count. Subsequent 2022 exchange-rate and yield paths, combined with constructed growth scenarios, produced lower client outcomes than the reference service although the selected services met the decision-time forecast benchmarks. These examples suggest that the approach could help make mandate choices and their consequences easier to examine. Professional and field studies could assess whether this improves oversight and client outcomes.

\end{abstract}


\section{Introduction}\label{sec:introduction}

A wealth manager delegating work to a system of AI agents asks it to make decisions from information filtered through two parties. The client may leave circumstances or preferences undisclosed; the manager may incompletely convey that account, its own objectives or the service conditions. Either omission can reflect strategic disclosure, uncertainty or imperfect articulation. The AI can have the least direct access to the private information on which its task depends, while being expected to serve both parties. The specification risk is that gaps are silently filled with learned defaults or unsupported assumptions about their circumstances and objectives, then carried into downstream decisions. Studies of multi-agent failures document action under incorrect assumptions, failures to seek clarification and incomplete verification \parencite{cemri2025fail}.

Agency theory addresses delegated authority under information asymmetry and potentially divergent interests \parencite{jensen1976theory}. A classical formulation advances the principal's interests while securing the agent's participation and desired behaviour \parencite{holmstrom1979moral}. Financial applications examine moral hazard in delegated portfolio management and incentives behind unsuitable sales \parencite{stoughton1993moral,inderst2009misselling}. Hierarchical agency supplies a basis for examining successive delegations \parencite{tirole1986hierarchies}. We propose to apply this framework to the client--manager--AI relationship: the client is the manager's principal, and the manager is both the client's agent and the AI system's principal. This structure locates responsibility for representing interests and transmitting information at each interface.

We propose to specify the AI's task through two explicit outcome functions: one for the client and one for the wealth manager. Client outcomes include return, liquidity and financial security; manager outcomes include service quality, resource use and commercial value. Their interests can coincide or conflict. The manager's interests shape the service whether or not the specification names them; naming them turns the trade-off into something a reviewer can read. Constrained joint maximisation over portfolio--workflow pairs makes the trade-off explicit, with legal duties, mandate requirements and evidence sufficiency determining admissibility. Weights and reference-service floors require justification and authorization. The intended engineering use is to translate this specification into agent instructions and workflow rules that preserve both evaluations, bind decisions to supported premises and identify when unresolved information or objective choices require human clarification. Each agent's assigned role must remain consistent with that shared specification.

The DACH setting makes the treatment of incomplete information consequential. In Switzerland, Articles~11--12 of the Financial Services Act (FinSA) require the provider to seek service-relevant client information. If that information is insufficient to assess appropriateness or suitability, Article~14(1) requires prior notice to the client that the provider cannot perform the assessment \parencite[arts.~11--14]{switzerland2018finsa}. For investment advice and portfolio management under EU suitability rules, Article~25(2) of the Markets in Financial Instruments Directive (MiFID~II) requires necessary client information, and Article~54(8) of Delegated Regulation~2017/565 bars recommendations when that information is missing \parencites[art.~25(2)]{europeanunion2014mifidii}[art.~54(8)]{europeancommission2017suitability}.

The manager must specify both which information the service requires and what an unresolved gap permits the system to do. A uniform hold policy adds an institutional restriction to the Swiss notification rule, while the EU suitability rule expressly restricts recommendations. Applicable mandate and other legal duties continue to govern the service. Jurisdiction and service type enter the decision specification because the same information gap can require different responses.

The research question is how nested client--manager--AI delegation can be translated into an inspectable specification for joint optimisation and clarification under incomplete information. The intended users are the manager responsible for the mandate and the engineer configuring the agents and workflow. The assessment distinguishes the adequacy of a transmitted specification, compliance with it and outcomes under subsequent market conditions. The simulation and executable constructions illustrate how the theory can be instantiated; the system architecture remains the builder's choice.

\section{Related Work}\label{sec:related-work}

Agency theory examines delegated authority under potentially divergent interests; portfolio-management and misselling models address effort, incentives and monitoring within the client--manager relationship \parencite{jensen1976theory,holmstrom1979moral,stoughton1993moral,inderst2009misselling}. Hierarchical agency provides the reference for a further delegation \parencite{tirole1986hierarchies}. Common agency studies an agent who serves several principals at once \parencite{bernheim1986common}; in our setting the two principals stand in sequence, and the second is the first one's agent. Work on automated advice examines what algorithmic portfolio services change for retail investors \parencite{dacunto2019roboadvising}.

PeterAI's publisher abstract describes personalised investment consultancy through specialised agents and an ambiguity-resolution pipeline; our comparison uses those reported architectural features \parencite{gadioli2026peterai}. Generative AI has also been modelled as an informational signal within principal-agent relationships \parencite{zhang2025genai}. Selective clarification and studies of multi-agent failures examine information requests and inherited errors \parencite{kuhn2023clam,cemri2025fail,xie2026spark}.

\textcite{zhuang2020misaligned} formalise incomplete objective specification as an AI principal--agent problem, analysing proxy objectives and interactive revision under stated assumptions. \textcite{ante2026specification} locates specification cost in persistent delegation losses with externally supplied objectives; the publisher abstract identifies incomplete specification as a financial-agent governance problem. A recent preprint by \textcite{sreenivas2026contracting} models contractual incentives when a delegated agent chooses its language model and inference effort.

A wealth-management benchmark evaluates paired assistant tasks with deterministic checkpoints \parencite{milsom2025benchmark}. The vendor-authored AI Advice Evidence Benchmark and associated Decision Assurance Envelope address investor context, mandate, suitability, policy, source evidence and human approval \parencite{srivastava2026evidence,srivastava2026envelope}; FiduciaryBench supplies source-linked US conduct and planning tasks \parencite{wealthschema2026fiduciary}. The Multi-Jurisdictional Legal Identity Assurance for Capability Gating model connects authority and supporting evidence to permission for an act, assigning mandate mechanics and aggregation to companion work \parencite{kurz2026identity}. Applying these approaches to wealth advice requires distinguishing client-authorized preferences from factual claims whose evidence requirements depend on the service.

Financial multicriteria decision support supplies an established methodological base \parencite{steuer2003finance}. Robust multi-objective optimization examines scalarization and dependent uncertainty \parencite{bokrantz2017robust}, while shield synthesis uses runtime monitoring to enforce specified properties \parencite{bloem2015shield}. These strands inform the distinction between selecting under declared evaluations and enforcing permission to act.

Austria's Securities Supervision Act 2018 (WAG~2018) and European Securities and Markets Authority (ESMA) guidance further specify assessment requirements \parencite{austria2018wag,esma2022suitability}. The EU General Data Protection Regulation (GDPR) and Swiss data-protection law govern processing \parencite{europeanunion2016gdpr,fdpic2025ai}. Institutional AI governance is addressed by the Swiss Financial Market Supervisory Authority (FINMA) and Germany's Federal Financial Supervisory Authority (BaFin) \parencite{finma2024ai,bafin2021bdai}. These sources distinguish information needed for the service from permitted collection and processing.

Existing rules retain responsibility for automated suitability assessment and require scrutiny of intended goals \parencites[art.~54(1)]{europeancommission2017suitability}[secs.~2.4 and 2.7]{finma2024ai}. ESMA guidance already specifies algorithm documentation, testing, controlled changes and internal sign-off, together with review of costlier equivalent products \parencite[paras.~90--95]{esma2022suitability}. Recent preprints connect financial-agent authority to mandate lineage, policy versions and approvals, and describe enforcement of delegated mandate chains \parencite{gong2026authority,racioppi2026mandato}. We formulate the delegated AI's task as constrained joint maximisation within the nested client--manager--AI relationship. The manager's dual role determines whose outcomes enter the objective, how legal and mandate duties restrict the choice set, and who is responsible for the specification transmitted to the AI.

\section{Contribution}\label{sec:contribution}

The contribution is a nested principal--agent formulation in which the delegated AI maximises a joint client--manager objective subject to binding constraints. The manager acts as the client's agent when agreeing the mandate and as the AI's principal when specifying its task. We connect these roles through separate client and manager outcome functions over portfolio--workflow pairs, with legal, mandate and evidence requirements determining admissibility before optimisation (\cref{sec:theory}). The weight places each decision between a client-first and a manager-first programme (Proposition~\ref{prop:limits}). Concession accounting separates benchmark restrictions from weighting; a proposed review screen requires policy review and assesses the total concession. Disclosure, translation and execution provide a decomposition of departures under stated counterfactual and tie-selection conventions. A simulation illustrates required-input recovery, numerical selection and version-bound approval. Exact constructions assess the review screen, while partition-based discretionary clarification is implemented for the restricted client selector.

Three research questions guide the assessment. First, how can client--manager--AI delegation be represented by a joint decision objective with binding client-protection, mandate and evidence constraints? Second, how do information omissions, representation choices and uncertainty assumptions affect admissibility, selection and the separate outcomes of both parties? Third, how can the specification be instantiated in agent instructions and workflow controls?

The formal specification and its finite client-selection instance address the first question. Analytical constructions and the portfolio--workflow simulation address the second, varying what each delegation transmits while retaining the original mandate for evaluation. The instruction comparison addresses the third, complemented by deterministic, single-agent and multi-agent implementations. Together these constructions demonstrate how declared objectives, service conditions and evidence requirements govern a decision.

\section{Methods}\label{sec:methods}

We used design science to develop the decision specification \parencite{hevner2004designscience,peffers2007dsrm}, combining analytical constructions, economic simulation and worked implementations for artificial evaluation \parencite{venable2016feds}. The formal results depend on the stated objectives, constraints and information structure; any executor preserving these rules can instantiate the specification. Exact-rational constructions isolated objective representation, benchmark eligibility, clarification and evidence transitions. Case policies supplied the assessment constraints informed by the legal mapping.

The economic simulation represented complete client circumstances, manager service terms and a declared mandate, then controlled what disclosure and specification transmitted. Four constructed families covered business-sale liquidity, retirement drawdown, currency liabilities and concentrated holdings. Three portfolios and three service workflows competed with a reference service. Separate interventions omitted client liabilities, omitted manager terms or changed the transmitted weight; faithful optimisation of each altered specification was evaluated against the original mandate. Simulated approval checked current premise support, a matching calculation, numerical constraints, reference floors and evidence version. The comparison held substantive mandate authorization fixed; the concession-review screen was assessed separately through exact constructions. Thus numerical execution and domain authorization have distinct roles in the demonstration.

Monthly cash flows included withdrawals, liabilities, currency holdings, zero-coupon bond valuation, fees, forced-sale costs and supervision. Client value combined expected wealth, a downside penalty and declared service value; manager value combined net fee income and service quality. Their scales and reference floors were fixed before selection. European Central Bank exchange rates and government yields supplied 144 monthly observations for 2014--2025 \parencite{ecb2026chfeur,ecb2026yieldcurves}. Growth returns, preferences, service conditions and three scenario probabilities were authored. Decision-time evaluations used the initial market observation; subsequent observed exchange rates and yields entered the outcome calculation with the same constructed growth scenarios.

Professional prose, explicit joint-objective instructions and explicit nested-delegation instructions received the same declared objective, evidence, questions, calculators, resource ceilings and simulated approval controls. Calculators returned all candidate outcomes for agent selection. December 2016 mandates supported development; the same four families at December 2021, with rescaled liabilities, spending and hourly costs, supplied four evaluation cases with eight information conditions each. These covered complete evidence, client or manager gaps, both gaps, translation conflict, refusal, correction after approval and equal-outcome controls. One run per treatment and condition yielded 96 assignments. All three instruction forms used the same model and runtime settings. Private states, answers and subsequent market paths stayed with the evaluator.

The retained studies tested supplied client-selection policies using deterministic, single-agent and multi-agent workflows. The structured pilot contained 108 assignments; each narrative comparison contained 126. Swiss Banking Ombudsman cases supplied authorization, retirement-depletion and investment-cash-flow mechanisms \parencite{ombudsman2023orders,ombudsman2020advice,ombudsman2019fund}; records and exposures were authored. The revised narrative interface combined typed extraction, clearer instructions and reusable arithmetic receipts, with re-evaluation on the same cases. These implementations illustrate the same decision concepts through alternative workflow arrangements.

Scoring distinguished fidelity to the original mandate, raw proposal defects, issuance, clarification burden, the declared joint optimum and separate economic outcomes. Current evidence and approval governed issuance. Independent cash-flow identities checked financial arithmetic; saved actions were replayed to verify decisions and evidence transitions. Source and model-response records retained configurations, prompts and usage. Conditions sharing a mandate were treated as dependent observations. Model expenditure and the constructed service costs were reported separately. The accompanying \emph{Supplementary Analytical and Resource Results} contains the detailed counterexamples, reference traces and resource tables.

The legal mapping concerns investment advice and portfolio management for private or retail clients, with transaction-specific advice and execution services supplying applicability contrasts.

Under Swiss FinSA, transaction-specific advice assesses knowledge and experience for the proposed instrument; portfolio-based advice and management also assess financial situation and objectives. Insufficient information triggers Article~14's client-information duty \parencite[arts.~11--14]{switzerland2018finsa}. German and Austrian advice and portfolio management use the suitability inputs in WpHG section~64(3), WAG~2018 section~56 and Article~54 of Delegated Regulation~2017/565, including loss-bearing capacity and risk tolerance \parencite{germany1994wphg64,austria2018wag,europeancommission2017suitability}. Article~54(8) conditions recommendations on obtaining the necessary information \parencite[art.~54(8)]{europeancommission2017suitability}. The artifact's uniform hold policy is a conservative design choice where the mapped Swiss provision specifies an information duty. The constructed liability amount is classified as required by the case policy.

Swiss execution or transmission services engage Article~13's notification rule; institutional-client treatment and professional-client waivers require the separate scope of Article~20 \parencite[arts.~13 and 20]{switzerland2018finsa}. EU appropriateness and qualifying execution-only services follow the conditions in MiFID~II Article~25(3)--(4), with professional-client assumptions addressed separately in the delegated rules \parencites[art.~25]{europeanunion2014mifidii}[arts.~54--56]{europeancommission2017suitability}. GDPR Articles~5, 6 and 9 govern processing principles and relevant grounds; Swiss processing is assessed under the Federal Act on Data Protection, including Articles~6 and 30--31 \parencite{europeanunion2016gdpr,switzerland2020fadp}. Client willingness, evidential adequacy and lawful processing receive separate policy decisions.

The legal analysis uses a cutoff of 30 September 2026: the FinSA consolidation of 1 March 2024, Regulation~2017/565 consolidation of 2 August 2022, current German section~64 text and Austrian section~56 effective 9 June 2022. MiFID~II Article~25 follows the consolidation of 6 June 2026. The German section~64 text is the version in force at the cutoff, cited without a provision-level effective date \parencite{switzerland2018finsa,europeancommission2017suitability,germany1994wphg64,austria2018wag,europeanunion2014mifidii}. FINMA Circular~2025/2, effective 1 January 2025, addresses service classification and knowledge-and-experience elicitation for providers within its scope \parencite[paras.~2, 4 and 13--14]{finma2025conduct}.

The constructed legal contrasts concern domestic retail services. Cross-border application depends on provider authorization, client category and the origin of the service request. Swiss territorial rules and German or Austrian market-access requirements can apply to the same relationship \parencites[art.~2]{switzerland2019finso}[sec.~15]{germany2021wpig15}[sec.~32]{germany1961kwg32}[secs.~21, 23--24]{austria2018wagcrossborder}. Exclusive client initiative has a fact-dependent scope, assessed from the service and solicitation history \parencite{esma2021reversesolicitation}. Market-access conditions, suitability requirements and the territorial scope of data-processing rules are assessed separately \parencites[art.~3]{europeanunion2016gdpr}[art.~3]{switzerland2020fadp}.

\section{From Classical Agency to Nested Delegation}\label{sec:theory}

Agency theory starts from a principal who cannot observe what the agent does or knows \parencite{jensen1976theory}. In the hidden-action formulation, the principal chooses a payment schedule $s$ and the effort $e$ it wants the agent to supply \parencite[pp.~75--76, eqs.~(1)--(3)]{holmstrom1979moral}:
\begin{equation}
 \begin{aligned}
 \max_{s\in\mathcal S,\,e\in\mathcal E}\quad&
 \mathbb E\!\left[u_P(Y-s(Y))\mid e\right]\\
 \text{subject to}\quad&
 \mathbb E\!\left[u_A(s(Y))\mid e\right]-c(e)\ge\bar u_A,\\
 &e\in\arg\max_{\tilde e\in\mathcal E}
 \left\{\mathbb E\!\left[u_A(s(Y))\mid\tilde e\right]-c(\tilde e)\right\}.
 \end{aligned}
 \label{eq:classical-agency}
\end{equation}
Here $Y$ is the observable outcome, $u_P$ and $u_A$ are the utilities of principal and agent, $c$ is the agent's effort cost, $\bar u_A$ its reservation utility, and $\mathcal S$, $\mathcal E$ are the admissible payment schedules and efforts. The first constraint secures the agent's participation. The second states that the agent picks the effort that suits it best under the offered payment.

Three features carry this model. The agent has interests of its own, expressed in $u_A$ and $c$. The agent holds the informational advantage, since it alone observes $e$. And the principal's instrument is the contract: it steers the agent through what it pays.

Wealth management with AI contains two delegations. The client delegates advisory work to the wealth manager, and the manager commissions and supervises the AI system:
\begin{equation}
 C=P_1\ \longrightarrow\ W=A_1=P_2\ \longrightarrow\ A=A_2.
 \label{eq:nested-delegation}
\end{equation}
$C$, $W$ and $A$ denote the client, the wealth manager and the AI system; $P_j$ and $A_j$ mark principal and agent in delegation $j$. Hierarchical agency has studied such chains among human parties \parencite{tirole1986hierarchies}. The first delegation is the familiar one: the manager knows more about products and about its own conduct than the client does, and its remuneration can pull against the client's interest \parencite{stoughton1993moral,inderst2009misselling}. The chain also differs from common agency, in which one agent serves several principals side by side \parencite{bernheim1986common}:
\begin{equation}
 \underbrace{P_1\ \longrightarrow\ A\ \longleftarrow\ P_2}_{\text{common agency}}
 \qquad\qquad
 \underbrace{P_1\ \longrightarrow\ (A_1=P_2)\ \longrightarrow\ A_2}_{\text{nested delegation}}
 \label{eq:common-nested}
\end{equation}
Under common agency the principals stand on equal footing and compete for the agent. In the nested chain the second principal is bound to the first, and this ranking is what later places admissibility before the objective.

\Cref{fig:nested-agency} shows the manager translating the client's mandate into AI instructions and receiving proposed advice with supporting premises. The dashed link denotes permitted client-context elicitation.

\input{3-assets/1-User/figures/nested-agency}

\emph{The instrument changes.} We model the AI as an executor of the supplied specification, with no separate utility, participation decision or strategic choice of effort. Computational and supervisory costs enter the manager's service evaluation. Under this assumption, the participation and incentive constraints of Equation~\eqref{eq:classical-agency} apply to the human service relationship; the second delegation is governed by the specification. The manager steers the system through the specification $P$ it hands down. $P$ contains an objective $F_P$, a set $\mathcal A_P$ of alternatives the system may recommend, and rules for the information it may rely on. Where the classical agent maximises its own utility under the contract, the system is asked to maximise the objective it was given, on the input $I_A$ it received:
\begin{equation}
 e\in\arg\max_{\tilde e\in\mathcal E}
 \left\{\mathbb E\!\left[u_A(s(Y))\mid\tilde e\right]-c(\tilde e)\right\}
 \qquad\text{becomes}\qquad
 a^X\in\arg\max_{a\in\mathcal A_P(I_A)}F_P(a\mid I_A).
 \label{eq:execution}
\end{equation}
The design variable moves from the payment schedule $s$ to the specification $P$ \parencite{zhuang2020misaligned}. A system that departs from Equation~\eqref{eq:execution} makes an execution error, which testing can find. The harder questions concern the two objects the system takes as given: what $P$ says, and what $I_A$ contains.

\emph{Information passes through two interfaces.} In Equation~\eqref{eq:classical-agency} the agent knows more than the principal. In the nested chain, the AI receives the client's private circumstances and the manager's service terms through disclosure and specification. Its access to that information depends on what those interfaces transmit and what supplementary evidence it can obtain. Let $\theta_C$ describe the client's circumstances and preferences and $\theta_W$ the manager's service objectives, costs and incentives. With a recorded mandate $M$ and jurisdiction and service conditions $J$, information reaches the AI in three steps:
\begin{equation}
 \begin{aligned}
 s_C &= D_C(\theta_C),\\
 P &= T_W(s_C,\theta_W,M,J,Z_W),\\
 I_A &= H_A(s_C,P,Z_A).
 \end{aligned}
 \label{eq:information-cascade}
\end{equation}
The client discloses $s_C$, the manager translates it together with its own terms into $P$, and the system receives the decision input $I_A$; $Z_W$ and $Z_A$ are supplementary evidence available at each stage. Each step can drop or distort content, deliberately \parencite{crawford1982strategic} or in good faith, and supplementary evidence can also add to what the system knows. Information omitted at an earlier step remains unavailable downstream unless clarification or supplementary evidence supplies it. The classical question is how the principal learns what the agent knows. Here the question is how what the principals know reaches the agent.

Disclosure can itself be a choice. A client whose true utility $U_C$ differs from the declared evaluation, for reasons of privacy, tax or attachment to a holding, selects what to disclose with the resulting recommendation in view:
\begin{equation}
 D_C(\theta_C)\in\arg\max_{s}\ U_C\bigl(a^X(s);\theta_C\bigr).
 \label{eq:strategic-disclosure}
\end{equation}
\textcite{crawford1982strategic} show that when the interests of sender and receiver differ, equilibrium messages are coarse: ranges of states share one message. In the notation of Equation~\eqref{eq:information-cascade}, several client states $\theta_C$ then map to the same $s_C$ and, without further evidence, to the same $I_A$. The following proposition states the limit this places on any system.

\begin{samepage}
\refstepcounter{theorem}\label{prop:indistinguishability}
Proposition~\thetheorem{} (indistinguishable client states). Fix the mandate, policy primitives and decision rule. Let two client states induce the same complete AI decision input $I_A$, but have disjoint sets $X_C^\star(\theta)$ and $X_C^\star(\theta')$ of fully informed client-optimal portfolios under those primitives. A rule selecting a portfolio solely from $I_A$ can guarantee a fully informed client optimum in at most one of these states.

\nopagebreak[4]
A deterministic rule receives the same input and selects the same portfolio in both states. That portfolio cannot belong to both disjoint optimal sets. A randomised rule has the same conditional output distribution in both states; probability-one optimality in each would require its support to lie in their empty intersection.
\par
\end{samepage}

The result concerns selection at the given input. A permitted question can distinguish the states; a hold preserves the unresolved decision. The proposition identifies an informational limit independently of the system's computational ability.

\emph{The conflict receives an explicit evaluation.} The manager is agent in the first delegation and principal in the second. As agent it owes the client loyalty; as principal it writes the system's objective. The proposed specification records the weight and scale assigned to its commercial interests together with the alternatives and reference service. Existing governance already requires documentation and review of automated decision rules \parencite[para.~90]{esma2022suitability}. Explicit dual evaluations add a measurable trade-off and a comparator against which a reviewer can examine each client concession.

We take the service arrangement and its remuneration as given. The mandate is also incomplete: circumstances change, so elicitation and specification have to be renewed \parencite{grossmanhart1986,hartmoore1990}. How client and manager respond strategically to a specification that both can read is a further layer of analysis, taken up in \cref{sec:limitations-future}.

The system decides on a pair $a=(x,\omega)$: a portfolio recommendation $x$ and the workflow $\omega$ that delivers it, drawn from a finite set $\mathcal A$. Let $E$ contain the accepted, currently valid case evidence from $I_A$. The declared evaluations $\widehat V_C(a\mid E)$ and $\widehat V_W(a\mid E)$ represent client and manager outcomes. Client criteria may include return, liquidity and security; manager criteria may include service quality, delivery cost and supervisory effort. Their definitions, scales and treatment of uncertainty are fixed before any comparison and remain subject to mandate review. An expectation needs an explicit belief model, a robust evaluation needs declared plausible states, and probabilities generated by a model need separate calibration before they serve as beliefs about a particular client.

Let $\mathcal H$ index the legal, risk, processing, mandate and independently justified service-viability constraints, including applicable client-interest duties, with residuals $g_h\le0$ when satisfied. Service viability records necessary delivery resources; preserving a reference profit is a separate benchmark choice. Let $G_{\mathrm{req}}(E,J,M)$ indicate that the information necessary for the service is established, and let $\Theta(E)$ be the non-empty set of states compatible with the evidence. Equation~\eqref{eq:joint-admissibility} admits the alternatives that pass the required-information gate and satisfy every constraint in every such state:
\begin{equation}
 \mathcal A_{\mathrm{adm}}(E,J,M)=\left\{a\in\mathcal A:
 \begin{array}{l}
 G_{\mathrm{req}}(E,J,M)=1,\\
 g_h(a;\theta,J,M)\le0\quad\forall h\in\mathcal H,\ \theta\in\Theta(E)
 \end{array}\right\}.
 \label{eq:joint-admissibility}
\end{equation}
For EU services, client-interest duties and restrictions on conflicting remuneration bound the manager's commercial objective \parencite[paras.~1 and 10]{esma2026mifid24}, alongside the rules on conflicts of interest \parencite[art.~23]{europeanunion2014mifidii}. For Swiss services, the corresponding bounds are the organisational duties on conflicts of interest and the rules on compensation from third parties \parencite[arts.~25--26]{switzerland2018finsa}, together with the agent's duties of loyalty and of accounting under mandate law \parencite[arts.~398 and 400]{switzerland1911co}. These rules enter as members of the constraint set, $\{h_{\mathrm{suit}},h_{\mathrm{conf}},h_{\mathrm{comp}}\}\subseteq\mathcal H$ for suitability, conflicts of interest and third-party compensation; the weighted comparison begins only where $g_h\le0$ holds for each of them. Missing required information or inconsistent evidence holds the recommendation.

A declared reference service $a_0$ gives each party a floor $b_r(E)=\widehat V_r(a_0\mid E)$, $r\in\{C,W\}$, under the same evidence and evaluation rules. The alternatives that are admissible and keep both parties at their reference level form the feasible set
\begin{equation}
 \mathcal A_{\mathrm{feas}}(E)=\left\{a\in\mathcal A_{\mathrm{adm}}(E,J,M):
 \widehat V_C(a\mid E)\ge b_C(E),\ \widehat V_W(a\mid E)\ge b_W(E)\right\}.
 \label{eq:feasible-joint}
\end{equation}

Within $\mathcal A_{\mathrm{feas}}(E)$ the first delegation can be written in the form of Equation~\eqref{eq:classical-agency}. A client who could choose the service itself would solve
\begin{equation}
 \max_{a\in\mathcal A_{\mathrm{feas}}(E)}\ \widehat V_C(a\mid E).
 \tag{C}\label{eq:client-programme}
\end{equation}
The client is the principal, and the manager floor plays the part of the participation constraint: the manager must do at least as well as with the reference service. The choice, however, lies with the manager. A manager who followed only its own interest would solve
\begin{equation}
 \max_{a\in\mathcal A_{\mathrm{feas}}(E)}\ \widehat V_W(a\mid E),
 \tag{W}\label{eq:manager-programme}
\end{equation}
where the client floor now takes the part of the participation constraint: the client keeps the mandate as long as it does at least as well as with the reference service. The distance between the solutions of \eqref{eq:client-programme} and \eqref{eq:manager-programme} is the agency problem of the first delegation, written for the service decision. A client who sees only the recommendation cannot tell which of the two programmes produced it.

What should a manager who takes both of its roles seriously write into $P$? Programme \eqref{eq:manager-programme} treats the second delegation as if the first did not exist. Programme \eqref{eq:client-programme} looks like the loyal choice, and it has a weakness of its own. The workflow $\omega$ uses the manager's staff time and supervisory capacity and determines the fee the client pays, so the manager's interests are part of the decision whether or not the objective names them. Under a client-only objective they enter elsewhere: in which alternatives are offered, which service tiers exist and how much review a case receives. Those choices are made outside the specification, where a reviewer sees no trade-off because none is written down.

We therefore propose that the specification states both evaluations and the weight between them. For a fixed $0<\lambda<1$, Equation~\eqref{eq:joint-objective} selects
\begin{equation}
 \begin{aligned}
 a^\star\in\arg\max_{a\in\mathcal A_{\mathrm{adm}}(E,J,M)}\quad&
 \lambda\widehat V_C(a\mid E)+(1-\lambda)\widehat V_W(a\mid E)\\
 \text{subject to}\quad&
 \widehat V_r(a\mid E)\ge b_r(E),\qquad r\in\{C,W\},
 \end{aligned}
 \label{eq:joint-objective}
\end{equation}
which is the maximisation of the weighted sum over $\mathcal A_{\mathrm{feas}}(E)$. The two programmes above are its limits.

\begin{samepage}
\refstepcounter{theorem}\label{prop:limits}
Proposition~\thetheorem{} (limiting programmes). Let $\mathcal A_{\mathrm{feas}}(E)$ be non-empty. Every solution of Equation~\eqref{eq:joint-objective} is Pareto-efficient in $\mathcal A_{\mathrm{feas}}(E)$ with respect to the two declared evaluations. There are thresholds $\lambda_W>0$ and $\lambda_C<1$ such that every solution solves \eqref{eq:manager-programme} when $\lambda<\lambda_W$ and solves \eqref{eq:client-programme} when $\lambda>\lambda_C$.

\nopagebreak[4]
An alternative in $\mathcal A_{\mathrm{feas}}(E)$ that weakly improved both evaluations of a solution and strictly improved one would have a strictly larger weighted sum, which contradicts optimality. For the thresholds, let $\delta_C>0$ be the smallest gap between the highest client value in $\mathcal A_{\mathrm{feas}}(E)$ and any lower client value, and let $R_W$ be the range of manager values in $\mathcal A_{\mathrm{feas}}(E)$. An alternative below the highest client value loses at least $\lambda\delta_C$ and gains at most $(1-\lambda)R_W$ against a client-best alternative, so it cannot be selected when $\lambda>\lambda_C=R_W/(R_W+\delta_C)$. The argument for $\lambda_W$ exchanges the two parties. If all alternatives share the highest client value, every solution solves \eqref{eq:client-programme} for any $\lambda$.
\par
\end{samepage}

The efficiency statement runs in one direction. On a finite set an efficient compromise can be selected at no weight: with client and manager values $(1,1/4)$, $(1/4,1)$ and $(3/5,3/5)$, the third pair scores $3/5$ at every $\lambda$ while one of the first two scores at least $5/8$. A mandate that wants such a compromise states it as a constraint.

The weight therefore has a plain reading: it places the decision between the manager's programme and the client's programme, and client priority is the case $\lambda>\lambda_C$. The proposal leaves the value of $\lambda$ to the mandate. It requires that the value be declared, justified against the mandate and authorized before the system acts on it.

The other elements have a reading in the nested structure as well. Admissibility comes first because the first delegation ranks above the second: legal duties and the mandate bound what the manager may want, so manager value competes only among alternatives that the client relationship already permits. The floors tie the decision to the reference service. The reservation utility $\bar u_A$ in Equation~\eqref{eq:classical-agency} is the value of an outside option; the manager floor $b_W$ is stricter, since it preserves the full managerial value of the reference service, and this choice needs its own justification in the mandate. An empty $\mathcal A_{\mathrm{feas}}(E)$ calls for revised alternatives or a held decision, and current human approval remains a condition of issuance.

Review requires the cost of both benchmark restrictions and weighting. Let $a_C$ solve \eqref{eq:client-programme}. Relative to that floor-feasible client optimum, the weight concession and manager gain are
\begin{equation}
 \Delta_C=\widehat V_C(a_C\mid E)-\widehat V_C(a^\star\mid E)\ \ge0,\qquad
 \Delta_W=\widehat V_W(a^\star\mid E)-\widehat V_W(a_C\mid E).
 \label{eq:concession}
\end{equation}
Optimality requires the manager gain to compensate for the weighted client concession:
\begin{equation}
 \lambda\widehat V_C(a^\star)+(1-\lambda)\widehat V_W(a^\star)\ \ge\
 \lambda\widehat V_C(a_C)+(1-\lambda)\widehat V_W(a_C)
 \quad\Longrightarrow\quad
 \Delta_W\ \ge\ \frac{\lambda}{1-\lambda}\,\Delta_C .
 \label{eq:exchange-rate}
\end{equation}
The rule accepts a client concession only when the manager gains at least $\lambda/(1-\lambda)$ times as much on the declared scales: nineteen times at $\lambda=0.95$, an equal amount at $\lambda=0.5$. The ratio is an exchange rate between the two evaluations, and it depends on how each is scaled. If client value is measured in units of $k_C$ euros and manager value in units of $k_W$ euros, one manager euro counts as much as $\varepsilon$ client euros, with
\begin{equation}
 \varepsilon(\lambda)=\frac{1-\lambda}{\lambda}\cdot\frac{k_C}{k_W}.
 \label{eq:euro-rate}
\end{equation}
A weight close to one can therefore still favour the manager when the manager scale is the finer one, and for $\varepsilon>1$ a pure transfer from client to manager raises the objective. The pair $(\Delta_C,\Delta_W)$ reports the effect of weighting within the floor-feasible set. To expose the effect of the floors themselves, let $a_H\in\arg\max_{a\in\mathcal A_{\mathrm{adm}}(E,J,M)}\widehat V_C(a\mid E)$ be a client-best hard-admissible and viable service, before reference floors. For non-empty feasible sets, total concession comprises benchmark and weighting effects:
\begin{equation}
 \begin{aligned}
 \Delta_C^{\mathrm{bench}}&=\widehat V_C(a_H\mid E)-\widehat V_C(a_C\mid E)\ge0,\\
 \Delta_C^{\mathrm{tot}}&=\widehat V_C(a_H\mid E)-\widehat V_C(a^\star\mid E)
 =\Delta_C^{\mathrm{bench}}+\Delta_C .
 \end{aligned}
 \label{eq:total-concession}
\end{equation}
A zero weight concession can therefore coexist with a positive benchmark concession. The proposed decision record reports both components, the total and the manager gain, with the comparator identities and policy version.

Delivery requirements enter $\mathcal H$. New evidence of a binding requirement triggers a revised admissible set and recomputation of $a_H$, $a_C$ and $a^\star$. This preserves comparison with the best remaining deliverable service. Let $Q(P,E)=1$ record completed independent policy review of the objectives, scales, weights, reference floors, service viability and applicable duties. Let $\beta(a^\star,a_H)$ be a supported, current incremental client benefit omitted from the declared evaluation, measured on its client scale; benefits already represented receive zero additional credit. We propose the following institutional screen for submitting a decision to substantive authorization:
\begin{equation}
 \mathsf{Ready}(a^\star,E,P)=
 Q(P,E)\ \land\
 \left[\Delta_C^{\mathrm{tot}}=0\ \lor\
 \beta(a^\star,a_H)\ge\Delta_C^{\mathrm{tot}}\right].
 \label{eq:authorization}
\end{equation}
For a positive total concession, absent or inadequate benefit evidence retains the decision for policy revision. An empty admissible or floor-feasible set holds selection. Passing the screen establishes readiness for domain assessment; the responsible reviewer determines compatibility with the mandate and applicable duties before authorization. The screen is a proposed review aid whose substantive adequacy requires validation. EU equivalent-service and instrument assessment additionally considers cost and complexity, with ESMA guidance addressing documented justification and review of costlier equivalent products \parencites[art.~54(9)]{europeancommission2017suitability}[paras.~91--95]{esma2022suitability}. Those obligations and the client-interest and conflict rules supply the legal basis for domain assessment \parencites[paras.~1 and 10]{esma2026mifid24}[arts.~25--26]{switzerland2018finsa}.

The chain in Equation~\eqref{eq:information-cascade} supplies a counterfactual decomposition of departures from an intended decision. Let $F^0(a)$ be the weighted sum of Equation~\eqref{eq:joint-objective} under the authorized mandate and complete information $(\theta_C,\theta_W)$. Fix a common deterministic rule for selecting among tied maximisers. Let $a^0$ be the choice made under those conditions, $a^D$ the choice the same rule makes from what the client disclosed, $a^T$ the choice it makes from the specification the manager transmitted, and $a^X$ the recommendation the system issues. For stages with a selected alternative, then
\begin{equation}
 F^0(a^0)-F^0(a^X)=
 \underbrace{F^0(a^0)-F^0(a^D)}_{\text{disclosure}}
 +\underbrace{F^0(a^D)-F^0(a^T)}_{\text{translation}}
 +\underbrace{F^0(a^T)-F^0(a^X)}_{\text{execution}}.
 \label{eq:interface-losses}
\end{equation}
The identity is elementary. Under these counterfactual conventions, its terms locate review responsibilities. The disclosure term concerns the client--manager interface and can motivate a permitted question; the translation term motivates review of the manager's specification; the execution term motivates conformance testing. A choice that the original mandate excludes is recorded as a violation at the interface where it arose. Holds remain separate dispositions. Single terms can offset one another; for originally feasible choices their sum is nonnegative.

A system that selects the transmitted optimum using the same tie rule makes $a^X=a^T$ and drives the execution term to zero. Membership in the argmax of Equation~\eqref{eq:execution} alone permits different tied choices. For example, transmitting $\lambda=2/3$ makes $B_1$ and $C_2$ tie in the construction in \cref{fig:joint-tradeoff}, while the authorized weight $3/4$ gives them values $7/8$ and $13/16$. Choosing $a^T=B_1$ and faithfully issuing $a^X=C_2$ assigns $1/16$ to execution under $F^0$. If the executed tie rule is unspecified, report the range of $F^0(a)-F^0(a^X)$ over transmitted maximisers; this example gives $[0,1/16]$. Regret measured under the transmitted objective is zero for every faithful maximiser. Responsibility for the specification remains with the manager under either reporting convention.

The specification fixes five things: two outcome functions that stay separate through every calculation, an authorized trade-off between them, admissibility checked before any comparison, an owner for every premise, and approval bound to the current evidence and policy version. The client confirms preferences, financial objectives and agreed service terms. The manager proposes delivery costs, commercial criteria, scales, the trade-off weight and a reference service. A domain reviewer independent of that commercial choice assesses the policy and each decision's compatibility with the client mandate and applicable duties, including both benchmark and weight concessions in $\Delta_C^{\mathrm{tot}}$. Domain review also establishes the service and evidence requirements that permission checks apply before comparison and issuance. The responsible manager authorizes the reviewed policy version. Client agreement records preferences and service scope; applicable legal duties remain binding constraints \parencite[paras.~1 and 10]{esma2026mifid24}.

In the restricted client selector, the manager specifies candidates, facet utilities, references and clarification thresholds; calculation tools apply Pareto filtering and assess the declared answer partitions. The agent identifies the premises supporting its proposal, while the evidence service tracks their validity and source ancestry. Human approval binds the candidate to the current evidence and policy version.

These responsibilities can be implemented through a single agent, several agents or a deterministic procedure. The analytical constructions and simulations in \cref{sec:results} examine the corresponding decision and evidence rules. Conformance establishes fidelity to the authorized specification; mandate and domain review assess its adequacy. A material change in objectives, alternatives, evidence criteria or permitted actions requires policy review and renewed assessment.

The finite client selector instantiates Equation~\eqref{eq:joint-objective} when the workflow and manager value are fixed and the floors exclude no feasible candidate. At an evidence state satisfying $G_{\mathrm{req}}=1$, let admissibility match $\mathcal F(E)$ and client value be a strictly increasing normalisation of the robust Nash product. The maximisers then coincide with Equation~\eqref{eq:nash-aggregate}; restricting candidates to $\mathcal P(E)$ gives the Pareto selector. \Cref{fig:recommendation-flow} and \cref{alg:decision} specify clarification and approval for this restricted case. The facets represent dimensions within the client outcome function. Discretionary clarification for the full joint objective, including a contingent sequence of questions, requires a further model of both parties' outcomes and answer probabilities or a declared robust question policy.

The evidence representation adapts the assertion/source distinction and reliance-event model developed for legal identity assurance \parencite{kurz2026identity}. Each premise carries its origin, supporting evidence, period of validity and use in the recommendation. Evidence sufficiency depends on the premise: a declared preference can be relevant on the client's authority, while a claim about an external liability may require corroboration.

Within the restricted client-selection procedure, facet agents represent client objectives through manager-approved utilities, reference values and evidence dependencies. The shared policy fixes their relation to the joint specification; agent decomposition must preserve that representation. Organisational roles supply hard legal, risk and data constraints. Evidence records retain source ancestry, so outputs derived from one premise share an origin and its subsequent invalidation \parencite{kurz2026identity}. Accepted factual evidence filters the enumerated joint-state set while preserving its declared dependencies.

Each context gap has a purpose, a decision relevance assessment and a permitted resolution route. The evidence officer assesses an answer's evidential adequacy and processing authorization. Acceptance filters the state set; refusal preserves the gap and its applicable route. Client willingness, evidential sufficiency and lawful processing remain separate decisions, including for special-category and third-party data \parencite[arts.~5--7, 9 and 14]{europeanunion2016gdpr}.

Let $X$ be a finite candidate set and $E$ the accepted, currently valid evidence. The policy supplies a finite joint-state set; compatible evidence selects the non-empty subset $\Theta(E)$. A conflict that leaves this subset empty holds the decision for resolution. For each active facet $i$, Equation~\eqref{eq:robust-utility} takes the lowest utility across the evidence-compatible states:
\begin{equation}
 \underline U_i(x;E)=\min_{\theta\in\Theta(E)}U_i(x;\theta).
 \label{eq:robust-utility}
\end{equation}
Here $U_i(x;\theta)$ is facet $i$'s utility in state $\theta$. Different facets can attain their minima in different states, so the aggregation protects componentwise lower bounds. Let $g_{\mathrm L},g_{\mathrm R},g_{\mathrm D}$ denote legal, risk and data-constraint residuals, with values at most zero indicating satisfaction. For fixed, state-independent reference values $d_i$, Equation~\eqref{eq:feasible} requires statewise admissibility and nonnegative facet surpluses:
\begin{equation}
 \mathcal F(E)=\left\{x\in X:
 \begin{array}{l}
 g_h(x;\theta)\le0\quad\forall h\in\{\mathrm L,\mathrm R,\mathrm D\},\ \theta\in\Theta(E),\\
 \underline U_i(x;E)\ge d_i\quad\forall i
 \end{array}\right\}.
 \label{eq:feasible}
\end{equation}
Each $d_i$ is the facet's declared reference utility. Evidence restriction can enlarge $\mathcal F(E)$ under fixed primitives, while required-information gates continue to govern issuance.

In the evaluated procedure, $k$ fixed facets represent client objectives. Equation~\eqref{eq:nash-aggregate} defines the provisional robust maximiser set $W(E)$ using the client Nash-product rule:
\begin{equation}
 W(E)=\arg\max_{x\in\mathcal F(E)}
 \prod_{i=1}^{k}\bigl(\underline U_i(x;E)-d_i\bigr).
 \label{eq:nash-aggregate}
\end{equation}
The product adapts the Nash criterion to finite multicriteria selection \parencite{nash1950bargaining}; $W(E)$ retains all exact ties. Facets are dimensions of the client's evaluation, with cardinal utilities, overlap policy and reference values fixed by the manager. A bargaining interpretation would require separate strategic and feasible-set assumptions. An uncertain reference requires a further convention, such as statewise surpluses. An empty $\mathcal F(E)$ holds the decision for human evidence resolution, revised alternatives or scope; the unfiltered rule retains zero-product ties for human adjudication.

The selection rule first removes candidates dominated across the current state/facet utility entries. Write $\mathcal P(E)$ for those $x\in\mathcal F(E)$ for which no $y\in\mathcal F(E)$ satisfies $U_i(y;\theta)\ge U_i(x;\theta)$ for every $i,\theta$, with strict inequality for at least one entry. Equation~\eqref{eq:pareto-selection} applies the Nash-product comparison to this statewise-undominated set:
\begin{equation}
 W^{\mathrm P}(E)=\arg\max_{x\in\mathcal P(E)}N_E(x),\qquad
 N_E(x)=\prod_i(\underline U_i(x;E)-d_i).
 \label{eq:pareto-selection}
\end{equation}
The score $N_E(x)$ is unchanged by the filter; $\mathcal P(E)$ restricts its domain using the declared state/facet utilities. This removes dominated zero-product choices and dominance concealed by equal robust scores. Remaining ties receive explicit human adjudication.

Let $\mathcal Q(E)$ index the unresolved inputs in the fixed declared schema whose possible recorded values induce nontrivial partitions. For input $q$, $\mathcal B_q(E)$ partitions $\Theta(E)$ by its possible recorded values. For each nonempty answer cell $B$, calculate $\mathcal F(B)$, $W^{\mathrm P}(B)$ and $N_B$ using the same primitives. With nonempty current feasibility, every current candidate remains feasible in each cell. For a current maximiser $x$, Equations~\eqref{eq:partition-money} and~\eqref{eq:partition-client} define separate monetary and client-criterion consequences of changing the selection after one declared input is resolved:
\begin{align}
 S_q^{\mathrm M}(x;E)&=\max_{\substack{B\in\mathcal B_q(E)\\\theta\in B}}
 \left[\ell(x,\theta)-\max_{y\in W^{\mathrm P}(B)}\ell(y,\theta)\right]_+,
 \label{eq:partition-money}\\
 S_q^{\mathrm C}(x;E)&=\max_{B\in\mathcal B_q(E)}
 \left[\max_{y\in W^{\mathrm P}(B)}N_B(y)-N_B(x)\right]_+.
 \label{eq:partition-client}
\end{align}
Here $\ell(x,\theta)$ is a separately declared euro loss and $[a]_+=\max(a,0)$. Both candidates in $S_q^{\mathrm C}$ are evaluated on the same answer cell $B$; a rise in the current candidate's own lower bound alone contributes no selection gain. The inner maximum in $S_q^{\mathrm M}$ requires improvement over the highest-loss answer-cell maximiser, so every resolved tie is retained. These are possible conditional gains under the declared criterion and tie convention. They relate to the value of information in decision analysis \parencite{howard1966information} as an upper bound. With answer probabilities $p(B)$ and a question cost $\kappa_q$ on the same scale, the expected net gain from asking $q$ before settling on $x$ is
\begin{equation}
 \mathrm{VOI}_q(x;E)=\sum_{B\in\mathcal B_q(E)}p(B)\Bigl[\max_{y\in W^{\mathrm P}(B)}N_B(y)-N_B(x)\Bigr]-\kappa_q
 \ \le\ S_q^{\mathrm C}(x;E)-\kappa_q.
 \label{eq:voi-bound}
\end{equation}
Each bracket is nonnegative, since $x$ stays feasible in every cell, and none exceeds the largest one, which is $S_q^{\mathrm C}$. A question that stays below the threshold $\tau_{\mathrm C}$ therefore also stays below it in expectation, whatever the probabilities. The converse needs the probabilities, which the present policy leaves undeclared.

The policy declares thresholds $\tau_{\mathrm M},\tau_{\mathrm C}\ge0$. Input $q$ blocks candidate $x$ when either consequence is positive and reaches its corresponding threshold. Equation~\eqref{eq:eligible} retains current maximisers whose consequences remain zero or below the corresponding threshold for every unresolved input:
\begin{equation}
 \mathcal R(E)=\left\{x\in W^{\mathrm P}(E):
 \begin{array}{l}
 \forall q\in\mathcal Q(E),\ r\in\{\mathrm M,\mathrm C\},\\
 S_q^r(x;E)=0\ \text{or}\ S_q^r(x;E)<\tau_r
 \end{array}\right\}.
 \label{eq:eligible}
\end{equation}
An empty partition collection leaves every member of $W^{\mathrm P}(E)$ eligible for review. The client threshold $\tau_{\mathrm C}$ uses Nash-product units. Positive affine transformations of utilities and references preserve selection; they preserve this test when $\tau_{\mathrm C}$ receives the corresponding product-scale multiplier. Monetary and client thresholds require separate materiality rationales.

The responsible manager, supported by domain review, must authorize both thresholds and their materiality rationale before execution. Monetary materiality is specified in euros for the mandate; client-criterion materiality requires comparisons of preference consequences on the declared utility scales. Zero thresholds block every positive consequence. The constructed checks stipulate threshold values to examine policy behaviour; practical calibration remains prospective.

Mandatory information, legal and provenance conditions govern every candidate. After these gates pass, a nonempty $\mathcal R(E)$ permits review of its members. If feasibility is nonempty but every current maximiser is blocked, the scheduler asks the first permitted unresolved input that blocks at least one current maximiser, within the fixed schema order and budget. It assesses declared partitions even when a question is unavailable or refused; those restrictions govern the lawful resolution route. Mandatory refusal, exhausted resources or unavailable consequential questions preserve a hold. A new accepted answer triggers reassessment, and a service-scope revision triggers a fresh applicability assessment. Necessary information remains a precondition for the relevant EU advisory service after human escalation \parencite{esma1762missinginfo}.

For comparison, we evaluate a global monetary clarification rule using the unfiltered $W(E)$ and singleton-state feasible and winner sets $F_\theta,W_\theta$. Equations~\eqref{eq:pivotality} and~\eqref{eq:consequence} define its global change indicator and monetary-consequence measure:
\begin{align}
 \Pi(E)&=\mathbf 1\{\exists\theta\in\Theta(E):(F_\theta,W_\theta)\ne(\mathcal F(E),W(E))\},\label{eq:pivotality}\\
 S(E)&=\max_{\substack{x\in W(E)\\\theta\in\Theta(E)}}
 \left[\ell(x,\theta)-\max_{y\in W_\theta}\ell(y,\theta)\right]_+.
 \label{eq:consequence}
\end{align}
Here $\mathbf 1$ is the indicator function, $F_\theta,W_\theta$ are the singleton-state feasible and winner sets, and $\bar s$ is the euro threshold. This comparator holds when $\Pi(E)=1$, $S(E)>0$ and $S(E)\ge\bar s$, alongside mandatory gates. Equations~\eqref{eq:partition-money}--\eqref{eq:eligible} instead assess each candidate over declared answer partitions and include client-criterion consequences. Both rules use the same finite-state representation and evidence criteria. The model experiments retain their executed client-selection policies; the partition rule is assessed through deterministic constructions.

\Cref{fig:recommendation-flow} connects the selection and clarification rules to the recommendation workflow. Mandatory checks precede comparison, consequential gaps prompt clarification or referral, and approval binds the selected candidate to the current evidence and policy version.

\input{3-assets/1-User/figures/recommendation-flow}

Human review receives the review-eligible maximisers, unresolved inputs, candidate-specific consequence diagnostics, gate outcomes and evidence/policy version. Approval binds a specific candidate to that version. Evidence acceptance, expiry, revocation and policy revision advance the version and clear approval. Human review is an architectural requirement; automated-decision duties require separate assessment under applicable data-protection law \parencites[art.~22]{europeanunion2016gdpr}{fdpic2026information}.

Context coverage $\kappa$ is the fraction of inputs in the manager-approved schema that satisfy their evidence criteria. The schema's completeness requires external validation. The record reports coverage, itemised gaps, gate outcomes and approval separately \parencite{kurz2026identity}.

Each transition records the version, actor role, source references and outcome. Revocation or expiry invalidates dependent premises transitively. Evidence acceptance, policy changes and approval are restricted to their declared roles. The reliance record distinguishes a premise's origin from the act of relying on it \parencite{kurz2026identity}. Role checks use trusted caller inputs. The event-time invariant assumes serialized transitions; deployment requires an atomic or version-conditioned issuance commit alongside identity, access, tamper-resistance and jurisdiction-specific storage controls.

The transition contract uses a decision version $v$ identifying current evidence and policy, a gate predicate $G_v$, and a recorded human approval $A_v(x)$. Equation~\eqref{eq:issuance} permits issuance only when the current gates pass and a review-eligible maximiser has human approval bound to the same version:
\begin{equation}
 \operatorname{issue}_v(x)\ \Longrightarrow\
 G_v=\operatorname{pass}\ \wedge\ x\in\mathcal R(E_v)\ \wedge\ A_v(x)=1.
 \label{eq:issuance}
\end{equation}
The predicate $G_v$ includes mandatory checks, consistent nonempty state support and robust feasibility; $A_v(x)$ binds human approval to candidate $x$ and version $v$. Issuance rechecks these conditions and current review eligibility.

\Cref{alg:decision} routes unresolved mandatory or consequential inputs to permitted questioning or a hold, and requires reassessment before renewed approval after evidence or policy changes.

\begingroup
\refstepcounter{algorithm}
\addcontentsline{loa}{algorithm}{\protect\numberline{\thealgorithm}Versioned recommendation procedure}
\begin{tcolorbox}[
  enhanced jigsaw,breakable,frame hidden,interior hidden,
  title style={draw=none,fill=none},coltitle=black,fonttitle=\normalfont\small,
  boxrule=0pt,boxsep=0pt,left=0pt,right=0pt,top=2pt,bottom=2pt,
  toptitle=2pt,bottomtitle=2pt,titlerule=0pt,
  before skip=\intextsep,after skip=\intextsep,
  borderline north={0.8pt}{0pt}{black},
  borderline south={0.4pt}{0pt}{black},
  overlay={\draw[line width=0.4pt] (title.south west)--(title.south east);},
  title={\swAlgorithmName~\thealgorithm: \textit{Versioned recommendation procedure for the restricted client selector}},
  title after break={\swAlgorithmName~\thealgorithm: \textit{Versioned recommendation procedure for the restricted client selector (continued from page~\pageref{alg:decision})}}
]
\label{alg:decision}
\begin{algorithmic}[1]
\Require Fixed policy, finite candidate and joint-state sets, evidence, question budget
\State Validate role, evidence dependencies and current policy version
\State Filter joint states by accepted valid evidence
\If{evidence is inconsistent or the joint-state set is empty}
  \State Hold for evidence resolution
\ElsIf{a mandatory gate fails}
  \State Request an unresolved permitted input within budget, or hold
\Else
  \State Compute $\mathcal F(E)$, $\mathcal P(E)$ and $W^{\mathrm P}(E)$ internally
  \If{$\mathcal F(E)=\varnothing$}
    \State Hold for human evidence resolution, revised alternatives or scope
  \Else
    \State Assess declared answer partitions and compute $\mathcal R(E)$
    \If{$\mathcal R(E)=\varnothing$}
      \State Request a consequential permitted input within budget, or hold
    \Else
      \State Present $\mathcal R(E)$, consequences, evidence, gaps and version for review
    \EndIf
  \EndIf
\EndIf
\State On evidence or policy change: invalidate approval and reassess
\State On approval: bind the selected review-eligible maximiser to the current version
\State On issuance: repeat checks and require current bound approval
\end{algorithmic}
\end{tcolorbox}
\endgroup

\begin{samepage}
\refstepcounter{theorem}\label{prop:gate}
Proposition~\thetheorem{} (gate preservation). Under the stated transition contract and trusted role boundary, every issuance event satisfies Equation~\eqref{eq:issuance} at its event time.

\nopagebreak[4]
The initial state carries no approval. Evidence and policy transitions clear approval and advance the version. Questioning and refusal leave issuance dependent on a subsequent successful assessment. The approval transition binds a selected review-eligible maximiser to the assessed version. The issuance transition re-evaluates the gates and checks membership in the current eligible set and approval/version equality. Each transition preserves the condition that issuance is possible only through these checks, establishing the event-time invariant by induction over the transition sequence.
\par
\end{samepage}

\begin{samepage}
\refstepcounter{theorem}\label{prop:expansion}
Proposition~\thetheorem{} (robust feasible-set expansion). Fix $X$, all utilities and constraints, the active facets and their reference values. For non-empty joint-state sets $\Theta'\subseteq\Theta$, the feasible sets defined by Equation~\eqref{eq:feasible} satisfy $\mathcal F(\Theta)\subseteq\mathcal F(\Theta')$.

\nopagebreak[4]
Any candidate feasible for every state in $\Theta$ satisfies the same constraints over its subset $\Theta'$. Taking a utility minimum over that subset weakly raises each lower bound. Every reference-value inequality satisfied under $\Theta$ remains satisfied under $\Theta'$, so each originally feasible candidate remains feasible.
\par
\end{samepage}

Recommendation permission also depends on mandatory information and versioned approval. An update can reveal a new liability, alter a constraint or invalidate a source; those changes require renewed assessment under the updated model. The inclusion concerns $\mathcal F$; Pareto membership, maximiser membership and review eligibility can change after a restriction of support.

\begin{samepage}
\refstepcounter{theorem}\label{prop:undominated}
Proposition~\thetheorem{} (undominated issuance). Under the issuance contract, an issued candidate is undominated among current robust-feasible candidates with respect to all declared state/facet utility entries.

\nopagebreak[4]
Equation~\eqref{eq:issuance} requires membership in $\mathcal R(E)$, which is a subset of $W^{\mathrm P}(E)$ and of $\mathcal P(E)$. Membership in $\mathcal P(E)$ excludes the stated dominance relation by definition, including when every feasible Nash product is zero.
\par
\end{samepage}

The guarantee depends on supplied utilities and states, whose adequacy for the mandate requires domain assessment.

\section{Results}\label{sec:results}

The analytical constructions establish how information, weights and reference floors affect selection. The portfolio--workflow simulation then measures those effects on the declared client and manager outcomes. Further results concern client-criterion sensitivity, clarification, evidence changes and execution by deterministic and model-based workflows.

The liability construction isolates how undisclosed client information changes selection and permission to issue advice. Client value in this example is the product of a return utility and a liquidity utility, the rule of Equation~\eqref{eq:nash-aggregate}. The synthetic client is an entrepreneur seeking advice on EUR~4 million of investment proceeds after a business sale. Three portfolio alternatives allocate different amounts to assets available within the next 12 months. The amount $q$ of a possible payment obligation within that horizon is the context variable. For this illustration, the service's suitability assessment requires that amount, and each candidate must provide liquid assets $L(x)\ge q$. All other legal, data and risk checks are assumed satisfied. \Cref{tab:worked-case} fixes the alternatives, with return and liquidity utility scores chosen solely to illustrate the mechanism and reference utilities $d_1=d_2=0$.

\begin{table}[htbp]
\centering
\caption{Synthetic alternatives and assumed facet utilities.}
\label{tab:worked-case}
\small
\begin{tabular*}{\textwidth}{@{\extracolsep{\fill}}lrrrrr@{}}
\toprule
Portfolio & Liquid assets & Other assets & Return utility & Liquidity utility & Nash product \\
 & EUR million & EUR million & $U_1$ & $U_2$ & $U_1U_2$ \\
\midrule
A & 0.4 & 3.6 & 9 & 2 & 18 \\
B & 1.2 & 2.8 & 6 & 6 & 36 \\
C & 2.0 & 2.0 & 3 & 9 & 27 \\
\bottomrule
\end{tabular*}
\end{table}

B maximises the product at an established obligation of EUR~0.2 million; C alone is feasible at EUR~1.8 million. With unresolved support $\{0.2,1.8\}$ million, C covers both amounts, while the mandatory-information gate holds issuance. An obligation above EUR~2 million requires revised alternatives or a continued hold. The unresolved input is identical whether the undisclosed obligation is EUR~0.2 million or EUR~1.8 million, while the fully informed winners are B and C. This instantiates Proposition~\ref{prop:indistinguishability}: robust C covers both amounts, but an accepted answer is needed to distinguish their client-optimal selections.

The joint construction adds the manager and the delivery workflow to this example, so that both declared evaluations take effect. Client values normalize the portfolio products in \cref{tab:worked-case} by 36. For an operational interpretation of the constructed workflows, $\omega_1$ uses general case review and $\omega_2$ uses template-supported review. We assume reusable templates reduce delivery burden for A and C, while B requires additional exception review under $\omega_2$. Both workflows are assumed to deliver the same assessed client outcome for a given portfolio. Manager values are stipulated as $\widehat V_W(x,\omega)=1-c(x,\omega)$, where $c$ is a normalized burden index: $(1/4,1/2,1/2)$ for A, B and C under $\omega_1$, and $(0,3/4,0)$ under $\omega_2$. Zero denotes the least burden on this illustrative scale. Empirical calibration of the burden index remains prospective.

The established EUR~0.2 million obligation makes all three portfolios financially feasible. The example stipulates that all six pairs pass the base admissibility conditions; domain review must establish which pairs can actually satisfy a mandate's client-interest duties. The client and manager floors are the values of reference service $(C,\omega_1)$, namely $(3/4,1/2)$. The floors retain $(B,\omega_1)$, $(C,\omega_1)$ and $(C,\omega_2)$.

\Cref{fig:joint-tradeoff} shows that $(B,\omega_1)$ and $(C,\omega_2)$ are the two efficient pairs meeting both floors; choosing between them requires a client--manager trade-off.

\input{3-assets/1-User/figures/joint-tradeoff}

At $\lambda=1/2$, Equation~\eqref{eq:joint-objective} selects $(C,\omega_2)$ with score $7/8$, compared with $3/4$ for $(B,\omega_1)$. At $\lambda=3/4$, $(B,\omega_1)$ wins with $7/8$, compared with $13/16$ for $(C,\omega_2)$. The two tie at $\lambda=2/3$, each scoring $5/6$. At the lower weight the selected pair preserves the client benchmark and improves the manager evaluation; at the higher weight it preserves the manager benchmark and improves the client evaluation. The weight makes an explicit choice along the declared trade-off, while the floors restrict permissible sacrifices.

In the terms of Proposition~\ref{prop:limits}, programme \eqref{eq:client-programme} selects $(B,\omega_1)$, programme \eqref{eq:manager-programme} selects $(C,\omega_2)$, and both thresholds equal $2/3$. At $\lambda=1/2$ the concession pair of Equation~\eqref{eq:concession} is $\Delta_C=1/4$ and $\Delta_W=1/2$, twice the minimum that Equation~\eqref{eq:exchange-rate} requires.

\Cref{fig:weight-sensitivity} shows selection moving from $(C,\omega_2)$ to $(B,\omega_1)$ as the client weight crosses $\lambda=2/3$, where the two tie.

\input{3-assets/1-User/figures/weight-sensitivity}

A capacity constraint excluding $(C,\omega_2)$ makes $(B,\omega_1)$ the winner at $\lambda=1/2$, even though the excluded alternative has the higher objective score. Missing required information holds every alternative. A further capacity construction admits only $(A,\omega_2)$ and $(B,\omega_2)$; each violates one floor, so the joint problem has no feasible solution and holds. The reference service remains the evaluation benchmark while capacity makes it unavailable. Two further capacity settings isolate each floor's effect on selection. With $(C,\omega_2)$ unavailable and $\lambda=1/4$, an objective-only comparison selects $(A,\omega_2)$ at $7/8$; the client floor excludes it, and the constrained rule selects $(B,\omega_1)$ at $5/8$. With only $(B,\omega_2)$ and $(C,\omega_1)$ available and $\lambda=3/4$, the objective-only winner is $(B,\omega_2)$ at $13/16$; the manager floor excludes it, and $(C,\omega_1)$ wins at $11/16$. The objective-only calculations isolate the floors' effects; the operative rule enforces both floors throughout. The eight configurations use constructed manager service evaluations; observed workflow expenditures enter the separate resource comparison below.

The two-pair capacity case gives a concrete review-screen test. Suppose the client confirms both service options and an independent cost assessment establishes their viability. With $(B,\omega_2)$ and $(C,\omega_1)$ hard-admissible, the client comparator before floors is $a_H=(B,\omega_2)$, while the manager floor leaves $a_C=a^\star=(C,\omega_1)$. Thus $\Delta_C=0$, $\Delta_C^{\mathrm{bench}}=1/4$ and $\Delta_C^{\mathrm{tot}}=1/4$. With zero omitted client benefit, Equation~\eqref{eq:authorization} returns the policy for revision even if policy review has been recorded. Choosing $(B,\omega_2)$ as the agreed reference instead gives floors $(1,1/4)$, selects that pair and reduces both concessions to zero.

A further construction isolates a delivery restriction. Let three services have client--manager values $(10,0)$, $(9,1)$ and $(1,20)$, zero floors and $\lambda=1/10$. Their scores are $1$, $9/5$ and $181/10$, so the third is selected. If a documented delivery requirement excludes the first service, it enters admissibility and the second becomes $a_H=a_C$. The remaining client concession is 8, which requires its own supported benefit assessment. These exact constructions also check that zero concession still requires policy review, that benefit evidence identifies the current comparator and policy, and that empty feasible sets hold. They implement the proposed review screen alongside the separate simulation.

The adversarial variations retain both party evaluations and both benchmark floors. Required-input omission produces a hold; accepted support refinement changes the robust admissible set; weight and reference translations change the selected pair. \Cref{tab:joint-stress} reports these effects using $B_1=(B,\omega_1)$, $C_1=(C,\omega_1)$ and $C_2=(C,\omega_2)$. The support-refinement condition uses a separate case policy: certified liability bounds satisfy its required-information gate, and the exact amount is optional. Except where varied explicitly, the reference is $C_1$, values are those of \cref{fig:joint-tradeoff}, and the weight is $1/2$. All 15 exact outcomes matched their hand-derived witnesses; the six common configurations matched the retained evaluator.

\begin{table}[htbp]
\centering
\caption{Controlled analytical stresses on the full joint objective.}
\label{tab:joint-stress}
\small
\begin{tabularx}{\textwidth}{@{}>{\raggedright\arraybackslash}p{0.20\textwidth}>{\raggedright\arraybackslash}X>{\raggedright\arraybackslash}p{0.21\textwidth}@{}}
\toprule
Variation & Controlled change & Joint outcome \\
\midrule
Required evidence & Established input becomes missing at $\lambda=3/4$. & $B_1$ to hold. \\
Liability support & Accepted support narrows from $\{0.2,1.8\}$ to $\{0.2\}$ million EUR at $\lambda=3/4$. & $C_2$ to $B_1$. \\
Weight translation & Authorized $\lambda=3/4$ is transmitted as $1/2$. & $B_1$ to $C_2$. \\
Reference translation & Authorized reference $C_1$ is transmitted as $B_1$. & $C_2$ to $B_1$. \\
Client-value constraint & Add a hard requirement $\widehat V_C\ge1$. & $C_2$ excluded; $B_1$ selected. \\
Workflow effect & Reduce $\widehat V_C(C_2)$ from $3/4$ to $1/2$, holding other values fixed. & $C_2$ fails its client floor; $B_1$ selected. \\
Affine representation & Transform $\widehat V_C$ to $2\widehat V_C+7$ and $\widehat V_W$ to $3\widehat V_W-2$, including benchmarks. & At $3/4$: $B_1,C_2$ tie. At $9/11$: $B_1$ restored. \\
\bottomrule
\end{tabularx}
\end{table}

The hard client-value requirement excludes $C_2$ before scalarisation even though it has the higher base weighted score. This checks precedence of an authorized constraint; institution-specific legal review supplies its substantive content. The workflow variation makes client value depend on delivery, relaxing the original example's invariance assumption. Under the affine transformation, changing $\lambda$ from $3/4$ to $9/11$ preserves every pairwise score difference up to a common positive factor $24/11$. Leaving the weight unchanged instead changes the effective policy. Inputs and preferences in these stress cases are authored.

The simulation illustrates the joint objective on mandates with financial content: both evaluations now depend on financial and service quantities. Constructed mandates, scales and weights make their effects explicit in a worked implementation. For each pair $a=(x,\omega)$ and scenario, define payment-adjusted economic wealth as $Y=A_T+H_{\mathrm{scheduled}}+L_{\mathrm{due}}-B_{\mathrm{unpaid}}$. Here $A_T$ is terminal financial wealth, $H_{\mathrm{scheduled}}$ scheduled household spending, $L_{\mathrm{due}}$ the liability valued at its payment-date exchange rate, and $B_{\mathrm{unpaid}}$ total unpaid bills, including service charges. Fees and transaction costs remain losses in this accounting measure; funding adequacy is enforced separately. Equation~\eqref{eq:simulation-values} gives the declared client and manager evaluations used in the joint objective:
\begin{align}
 \widehat V_C(a)&=\frac{\mathbb E[Y(a)]-A_0-\rho_C\bigl(\mathbb E[Y(a)]-\min_{\theta\in\Theta}Y(a;\theta)\bigr)+v_Cs_\omega}{0.1A_0},\nonumber\\
 \widehat V_W(a)&=\frac{\pi(a)}{0.01A_0}+0.1s_\omega.
 \label{eq:simulation-values}
\end{align}
Here $A_0$ is initial wealth, $\rho_C$ the client's declared downside weight, $v_C$ its euro value per unit of service quality $s_\omega$, and $\pi(a)$ manager fee income less personnel and fixed costs. Expectations use three authored scenarios with probabilities $0.25$, $0.50$ and $0.25$; $\Theta$ is their support. Both reference floors use these same definitions. The client and manager scales are fixed at 10\% and 1\% of initial wealth. Hard constraints separately enforce liquidity reserves, loss capacity, issuer concentration, funded obligations and supervisory capacity.

The four complete evaluation mandates admit ten, five, ten and seven pairs, respectively, after applying hard constraints and both reference floors. Each retains at least two undominated client--manager trade-offs. Both floors admit every hard-feasible pair in all eight original and corrected states; their restrictive effects are assessed in the analytical constructions. At client weights $0.95$, $0.90$, $0.85$ and $0.98$, the business-sale, retirement, currency-liability and concentration mandates select growth/standard, balanced/advisory, growth/specialist and growth/advisory services. These names identify portfolio and delivery choices; the instruction treatments use the same agent architecture.

\Cref{tab:concession} gives the price of these weights by comparing each selection with the solution of programme \eqref{eq:client-programme} over the same feasible set. With client value scaled to $k_C=0.1A_0$ and manager value to $k_W=0.01A_0$, Equation~\eqref{eq:euro-rate} becomes $\varepsilon=10(1-\lambda)/\lambda$. In the two business-sale states the selected pair is also the client's best. In the other six states the rule selects the advisory or specialist workflow where the client-best pair uses the standard workflow, and in two of the four mandates a manager euro counts for more than a client euro.

\begin{table}[htbp]
\centering
\caption{Client concession and manager gain against the client-best pair at the declared weights.}
\label{tab:concession}
\small
\begin{tabularx}{\textwidth}{@{}>{\raggedright\arraybackslash}Xrr>{\raggedright\arraybackslash}p{0.27\textwidth}rr@{}}
\toprule
Mandate & $\lambda$ & $\varepsilon$ & Selected and client-best workflow & $\Delta_C$ (EUR) & Manager gain (EUR) \\
\midrule
Business sale & 0.95 & 0.53 & Standard; identical & 0 & 0 \\
Retirement drawdown & 0.90 & 1.11 & Advisory; standard & 1,190 & 3,062 \\
Currency liability & 0.85 & 1.76 & Specialist; standard & 2,264 & 10,381 \\
Concentrated holdings & 0.98 & 0.20 & Advisory; standard & 1,178 & 6,656 \\
\bottomrule
\end{tabularx}
{\raggedright\fignote{Original and corrected liability states of a mandate give the same differences. $\Delta_C$ is the difference in the declared client criterion of Equation~\eqref{eq:simulation-values}, converted to euros; the manager gain is the difference in net income $\pi$. Clarification charges are zero in both selections. The portfolio is the same within each comparison.}}
\end{table}

The reference service holds cash at a fee of 0.4\%. Every hard-feasible pair meets both of its floors, so $\mathcal A_{\mathrm{feas}}(E)$ equals the hard-feasible set in these mandates and the floors restrict nothing here. Consequently $\Delta_C^{\mathrm{bench}}=0$ here, and \cref{tab:concession} reports the total as well as the weight concession. The six positive concessions identify the additional client-benefit evidence that Equation~\eqref{eq:authorization} would require in a substantive mandate review; the simulation implements the declared weighted selection.

\Cref{fig:nested-tradeoff} shows the client--manager trade-off for the business-sale mandate. Lowering the transmitted client weight from $0.95$ to $0.35$ redirects the choice from growth/standard to growth/specialist, while both remain admissible.

\input{3-assets/1-User/figures/nested-tradeoff}

Eight evaluation states combine the original and corrected liability for each mandate. Faithful optimisation of the complete specifications preserves all eight original decisions. Setting the omitted liability to zero produces two recommendations that violate the original liquidity requirement; replacing omitted manager terms with zero hourly cost and 100 available supervisory hours produces two violations of supervision capacity. Transmitting a client weight of $0.35$ changes four optimal choices while each selected pair remains admissible under the original constraints. These are deterministic interventions in specification content. The optimiser follows the received objective in every condition; the original mandate supplies the reference for assessing fidelity. In the terms of Equation~\eqref{eq:interface-losses}, the omitted liability acts on the disclosure term, the omitted manager terms and the mistranslated weight act on the translation term, and the execution term is zero under the common deterministic tie rule used in these controls. \Cref{fig:nested-specification} separates these constraint failures from admissible changes in the authorized trade-off.

\input{3-assets/1-User/figures/nested-specification}

The instruction comparison keeps the declared decision content fixed across three forms. Each reaches all 32 prescribed terminal decisions: 28 issuances and four justified holds. Every issuance preserves the declared numerical mandate, current simulated approval and the joint optimum; unsupported proposals and violations of the encoded constraints are zero. Regret under the declared objective is zero for all 84 issuances, alongside 12 required holds. The equal-output controls assign identical original values to every tied choice, so tie selection leaves the attribution unchanged in these records. Professional prose and nested-delegation instructions each use 20 questions; the explicit joint-objective treatment uses 26. Its six additional questions incur EUR~450 in nominal client charges and EUR~672 in manager labour cost. \Cref{tab:nested-instructions} separates these constructed clarification costs from metered model expenditure.

\begin{table}[htbp]
\centering
\caption{Execution of the joint objective under three instruction forms.}
\label{tab:nested-instructions}
\begin{tabularx}{\textwidth}{@{}>{\raggedright\arraybackslash}Xrrr@{}}
\toprule
Measure & Professional prose & Joint objective & Nested delegation \\
\midrule
Correct terminal decisions & 32/32 & 32/32 & 32/32 \\
Issued recommendations / required holds & 28/4 & 28/4 & 28/4 \\
Unsafe raw proposals & 0 & 0 & 0 \\
Issued numerical mandate or constraint violations & 0 & 0 & 0 \\
Questions & 20 & 26 & 20 \\
Client question charges (EUR) & 1,500 & 1,950 & 1,500 \\
Manager question labour (EUR) & 2,352 & 3,024 & 2,352 \\
Reported model cost (USD) & 6.3146 & 7.3317 & 6.3162 \\
\bottomrule
\end{tabularx}
{\raggedright\fignote{Four constructed mandates, eight dependent information conditions per mandate, one repetition. Objectives, evidence access, arithmetic and approval controls are shared. Each question incurs a constructed EUR~75 client charge and half an hour of manager labour at the mandate-specific rate. All costs include holds; model expenditure is metered separately.}}
\end{table}

The comparison contains 96 model episodes; reconstruction reproduces their decisions, evidence transitions and economic outcomes. A public-input reference policy reaches all 64 development and evaluation conditions. The common runtime supplies authoritative source ownership and enforces supported premises, admissibility and current approval. The three instruction forms illustrate equivalent execution of the shared decision rule within those controls.

The subsequent market-path calculation keeps each selected pair fixed and replaces the declared exchange-rate and yield paths with the 2022 observations, retaining the three constructed growth scenarios and their probabilities. The client outcome is $Y$ plus declared service value under the accounting definition above. Each reference service bears the same clarification expenditure as its corresponding episode. Averaging the eight distinct original and corrected mandate states equally gives a client measure EUR~77,967 below the reference service and manager net income EUR~8,816 above it in every treatment. Manager net income follows from fee income less personnel and fixed costs:
\begin{equation}
 \pi(a)=f_\omega A_0-c_W\Bigl(\tau_\omega+\tfrac{n}{2}\Bigr)-500,
 \label{eq:manager-income}
\end{equation}
with workflow fee rate $f_\omega$, hourly personnel cost $c_W$, workflow hours $\tau_\omega$, $n$ questions at half an hour each and a fixed cost of EUR~500. No term depends on the market path, so the manager figure would be the same in any year; the client figure alone reflects 2022. Equal-output controls and repeated information conditions enter the decision counts; the economic mean uses each substantive state once. Question-cost totals report clarification burden separately. All obligations are funded. These dependent conditional outcomes differ from the risk-adjusted decision-time client criterion.

\Cref{fig:nested-outcomes} compares forecast and subsequent-path client outcomes using the same euro measure, relative to the same reference service. The five-year government yield rises from approximately $-0.48\%$ at the decision date to $2.45\%$ at the end of 2022, exceeding the stipulated adverse-scenario increase of 1.5 percentage points. The reference service holds cash, whose accrual proxy also rises. This identifies a consequential mismatch between the declared market scenarios and subsequent observations. The outcome calculation preserves the constructed growth assumptions, so interpretation concerns that specified combination of historical and authored paths.

\input{3-assets/1-User/figures/nested-outcomes}

The representation checks in \cref{tab:sensitivity} hold the candidate set fixed while changing the aggregation specification. Duplicating the liquidity facet changes the complete-context products from $(18,36,27)$ to $(36,216,243)$ and selects C. Setting reference values to $(0,4)$ leaves B and C with products 12 and 15. Positive affine utility transformations with correspondingly transformed references preserve the winner: each facet surplus is multiplied by a positive constant, so every candidate product receives the same positive multiplier. For an exact duplicate facet, splitting an existing exponent between its copies preserves that facet's product contribution; the exponent policy remains part of the declared representation.

\begin{table}[htbp]
\centering
\caption{Constructed sensitivity checks for the allocation example.}
\label{tab:sensitivity}
\small
\begin{tabularx}{\textwidth}{@{}Xrrrl@{}}
\toprule
Specification & A & B & C & Maximiser \\
\midrule
Original product & 18 & 36 & 27 & B \\
Liquidity represented twice & 36 & 216 & 243 & C \\
Reference values $(0,4)$ & Infeasible & 12 & 15 & C \\
Equal-weight utility sum & 11 & 12 & 12 & B, C \\
\bottomrule
\end{tabularx}
\end{table}

Taking the original two-facet specification as the declared representation of the mandate, the duplication in \cref{tab:sensitivity} changes its implemented weighting while retaining every original utility value and financial constraint. The resulting C choice satisfies the duplicated policy; the original policy selects B. The construction separates execution of a specified aggregation rule from preservation of the declared mandate representation. Which representation reflects a particular client's preferences requires evidence beyond the stipulated utilities.

All 18 constructed aggregation cases retained two to four robust-feasible candidates. Of the 17 equal-weight comparisons, seven had identical winner sets, six had overlapping but unequal sets and four had disjoint sets. Fifteen cases had conflicting rankings among robust-feasible alternatives; these accounted for five identical sets, six overlapping sets and all four disjoint sets. The separately weighted sensitivity also produced disjoint winners. These counts describe related constructions selected to examine policy behaviour.

A strict disagreement occurs with normalized utilities A=$(1,0.3)$, B=$(0.6,0.6)$ and C=$(0.3,0.9)$, zero references and common feasibility. Nash products of $0.30$, $0.36$ and $0.27$ select B. Equal-weight sums of $0.65$, $0.60$ and $0.60$ select A. The shared constraints admit both selections; their relative desirability depends on the client criterion used to assess the chosen scalarisation. Exact witnesses and reference checks reproduced all 18 Nash calculations.

A further construction isolates the uncertainty objective. With fixed zero reference values, candidate X has utility vectors $(1,9)$ and $(9,1)$ in two joint states; candidate Y has $(2,2)$ in both. Both candidates satisfy all constraints. The product of componentwise minima scores X at 1 and Y at 4, selecting Y. The worst joint-state product, $\min_{\theta}\prod_i(U_i(x;\theta)-d_i)$, scores X at 9 and Y at 4, selecting X. The two policies protect different quantities even though the joint states are identical. If X incurs zero monetary loss and Y incurs EUR~100 in each state, the unresolved robust choice has EUR~100 regret. Comparing only the singleton winners would miss this clarification opportunity because both states select X. Equation~\eqref{eq:pivotality} also compares each resolved choice with the unresolved choice, detecting the difference; the executed case asks at a EUR~10 threshold.

In the constructed comparisons, statewise Pareto filtering removed a dominated candidate whose equal robust score had blocked an eligible alternative, and excluded a dominated choice when all Nash products were zero. Assessing answer cells detected a partial answer that changed the selected portfolio although the unresolved and all singleton choices agreed. Candidate-specific eligibility also allowed an unblocked maximiser to proceed while a tied peer awaited clarification. The full numerical witnesses and comparator traces are reported in the supplementary analytical results.

The partial-answer case shows why the answer partition matters. A declared input separates two of three states from the third. In that two-state cell, changing from A to B increases the client criterion by $4-1=3$ and reduces monetary loss by EUR~100. The rule asks at a client threshold of 1 or a monetary threshold of EUR~10. A further answer identifying either remaining state changes the preferred candidate back to A, with a client-criterion gain of $9-4=5$. Selection can therefore reverse as evidence arrives under the componentwise robust criterion.

The client-criterion test also detects consequences at equal monetary loss. Candidate A has utilities $(1,100)$ and $(100,1)$ in two states; B has $(2,2)$ in both, with zero references and zero losses. The unresolved products are 1 and 4, while each singleton selects A with product 100. Resolving the state yields a client-criterion gain of $100-4=96$, triggering a question at a threshold of at most 96. The monetary comparator leaves B eligible for review.

The diagnostics also left consequential information unresolved in two constructions. With flat Nash products and equal monetary losses, both diagnostics were zero although each answer changed which candidate was undominated. With two optional binary inputs, each individual diagnostic was zero while joint resolution changed the winner from B, with product 4, to A, with product 100. Zero individual consequences therefore leave both dominance changes and joint materiality possible. The supplementary witnesses give the utilities and answer partitions.

When an optional answer would restore an empty robust-feasible set, the procedure held for human evidence resolution. Its diagnostics require a feasible current maximiser, so the autonomous question request was rejected. Externally accepted evidence for either answer restored review of the corresponding candidate.

The finite procedure passed 18 checks comprising 26 constructed traces and 14 issuance events. Evidence withdrawal cleared approval even when an irrelevant source or a second independent corroborating source left mandatory support intact. Withdrawal of the sole required source held the decision until accepted replacement evidence restored support and a new approval authorized issuance. Expiry produced the same reassessment under the explicitly advanced logical clock; earlier issuance events remained in the record. Role checks rejected unauthorized evidence, policy and approval operations at the trusted application programming interface (API) boundary. These results establish conformance to the supplied evidence and feasibility labels.

Applying two Swiss service policies to the same allocation states changed permission to proceed. Knowledge and experience and all other relevant service inputs are stipulated as established. The portfolio-based policy classifies the material liability as required financial information; the transaction-specific policy limits required inputs to knowledge and experience. Both retain robust candidate C with zero unmet-liquidity loss, while the portfolio-based policy asks and then holds after refusal and the transaction-specific policy permits review. The difference follows from the declared applicability mapping and conservative institutional hold, with the financial state set held fixed.

\Cref{tab:case-states} contrasts the mandatory amount with an optional communication-format preference that leaves utilities and constraints unchanged.

\begin{table}[htbp]
\centering
\caption{Decision traces under four synthetic information conditions.}
\label{tab:case-states}
\small
\begin{tabularx}{\textwidth}{@{}>{\raggedright\arraybackslash}p{0.21\textwidth} >{\raggedright\arraybackslash}X >{\raggedright\arraybackslash}X@{}}
\toprule
Condition & Available context & Model outcome \\
\midrule
Complete & Required information established; $q=0.2$ million. & B proceeds to human review. \\
Missing pivotal input & Required amount unresolved between 0.2 and 1.8 million. & Ask and hold; reassess after an adequate answer. \\
Missing optional context & $q=0.2$ million established; communication-format preference absent. & B proceeds to review; optional gap remains recorded. \\
Required disclosure withheld & Client declines to supply the necessary obligation information. & Hold and refer to the advisor; mandatory checks remain active. \\
\bottomrule
\end{tabularx}
\end{table}

A currency mismatch illustrates how a changed valuation can invalidate an earlier liquidity assessment. Assume liquid euro assets of EUR~1.2 million and an unhedged CHF~1.4 million payment obligation, with interest, taxes and execution costs omitted. Each date values the same assumed exposure separately. If $e_t$ is the observed number of Swiss francs per euro, its euro cost is $q_t=1{,}400{,}000/e_t$, and the liquidity shortfall is $\max(0,q_t-1{,}200{,}000)$. \Cref{tab:fx-replay} reports the calculation using observed ECB reference rates \parencite{ecb2026chfeur}.

\begin{table}[htbp]
\centering
\caption{Constructed CHF liability valued at observed ECB rates.}
\label{tab:fx-replay}
\small
\begin{tabular*}{\textwidth}{@{\extracolsep{\fill}}lrrr@{}}
\toprule
Date & CHF per EUR & Liability in EUR & Shortfall in EUR \\
\midrule
14 January 2015 & 1.2010 & 1,165,695.25 & 0.00 \\
15 January 2015 & 1.0280 & 1,361,867.70 & 161,867.70 \\
16 January 2015 & 1.0128 & 1,382,306.48 & 182,306.48 \\
\bottomrule
\end{tabular*}
\end{table}

Using the 14 January valuation on either subsequent date would retain an apparent surplus despite the shortfall at the current rate.

The workflow comparisons test whether supplied instructions produce the specified questions, holds and recommendations from partial evidence, while recording delivery resources. All workflows reached the 36 reference decisions in the structured pilot (\cref{tab:pilot-outcomes}), with consistent selections across repetitions. Each made 18 question attempts. Unsafe proposals, issued violations, unnecessary holds and schema failures were zero.

\begin{table}[htbp]
\centering
\caption{Development-pilot outcomes and resources across 36 records per workflow.}
\label{tab:pilot-outcomes}
\small
\begin{tabularx}{\textwidth}{@{}>{\raggedright\arraybackslash}X*{3}{>{\raggedleft\arraybackslash}p{0.17\textwidth}}@{}}
\toprule
Measure & Deterministic & Single agent & Multi-agent \\
\midrule
Correct terminal decisions & 36/36 & 36/36 & 36/36 \\
Correct permitted recommendations & 24 & 24 & 24 \\
Required holds & 12 & 12 & 12 \\
Unsafe raw proposals / proposals & 0/24 & 0/24 & 0/24 \\
Issued violations / issued outputs & 0/24 & 0/24 & 0/24 \\
Reported model cost (USD) & 0.0000 & 0.9758 & 3.1034 \\
\bottomrule
\end{tabularx}
\end{table}

The multi-agent workflow incurred 3.18 times the single agent's reported model cost. The deterministic workflow used the shared calculator and incurred zero model charges. Implementation, local computation and professional review costs fall outside this accounting boundary.

The retirement construction sets assets at EUR~300,000 and annual pension income at EUR~24,000. Essential expenditure is EUR~5,000 per month and a separate EUR~60,000 payment falls within a three-year reserve horizon. Equation~\eqref{eq:narrative-reserve} gives the required liquid reserve $R$:
\begin{equation}
 R=\max(b-p,0)h+o
  =\max(60{,}000-24{,}000,0)\cdot3+60{,}000
  =168{,}000\;\text{EUR}.
 \label{eq:narrative-reserve}
\end{equation}
Here $b,p$ are annual expenditure and pension income, $h$ is the horizon in years and $o$ the separate obligation. Only R180, the EUR~180,000 reserve alternative, covers $R$; the other reserves are EUR~60,000 and EUR~120,000. A competing questionnaire records EUR~3,000 monthly expenditure, implying a EUR~96,000 requirement and a different feasible set. The signed reconciliation establishes which figure governs. Expenditure and the separate obligation are required fields in the case policy.

The second construction separates the investment's price change from its complete cash flows. Equation~\eqref{eq:narrative-cashflows} calculates available assets $A$ and the net investment result $P$:
\begin{align}
 A &=s+d-f=82{,}000+4\cdot5{,}500-(2{,}000+4{,}000)=98{,}000,
 \nonumber\\
 P &=A-i=98{,}000-100{,}000=-2{,}000.
 \label{eq:narrative-cashflows}
\end{align}
Here $s$ denotes sale proceeds, $d$ retained distributions, $f$ separately debited fees and taxes, and $i$ purchase cost, all in EUR. A near-term EUR~90,000 liability makes the EUR~80,000 reserve insufficient, while EUR~100,000 exceeds available assets $A$. R90, the EUR~90,000 alternative, satisfies both bounds. The EUR~2,000 loss records the investment result $P$; feasibility uses available assets. All distributions remain available and the separate charges exhaust the stipulated fees and taxes.

\Cref{tab:narrative-outcomes} gives the four issuance and three hold conditions in each narrative family. The withdrawal grade requires valid initial proposal and approval before evidence loss; a blanket initial hold fails that sequence.

\begin{table}[htbp]
\centering
\caption{Information variants and conformance outcomes in both narrative families.}
\label{tab:narrative-outcomes}
\small
\begin{tabularx}{\textwidth}{@{}>{\raggedright\arraybackslash}p{0.24\textwidth}>{\raggedright\arraybackslash}X>{\raggedright\arraybackslash}p{0.16\textwidth}@{}}
\toprule
Information variant & Required evidence transition & Terminal outcome \\
\midrule
Complete evidence & Current records support calculation and approval. & Issue \\
Two jointly missing inputs & Obtain both required records; either adequate question order is accepted. & Issue \\
Refusal & A required input remains unresolved after the client declines. & Hold \\
Resolvable conflict & Signed reconciliation identifies the superseded assertion. & Issue \\
Unresolved conflict & The reconciliation route returns unavailable. & Hold \\
Withdrawal after approval & Evaluator invalidates the source dependencies and approval; workflow responds to the changed state. & Hold \\
Missing presentation preference & Financial inputs remain complete and unchanged. & Issue \\
\bottomrule
\end{tabularx}
\end{table}

Under fixed responses, all three interfaces reached the required outcomes in all 14 cases; reconstruction reproduced those outcomes.

The initial narrative comparison yielded 42, 30 and 22 correct decisions for deterministic, single-agent and multi-agent workflows, respectively (\cref{tab:narrative-live}). Model-workflow rates were 71.4\% and 52.4\%; correct recommendations numbered 14 and seven out of 24 issuance cases. Each workflow asked 30 relevant questions and zero redundant questions.

\input{3-assets/1-User/tables/narrative-live-results}

The calculator accepted exactly the required extracted fields and read remaining quantities from the public fixed inputs. Additional fields caused all rejected calculator calls, while repeated successful calculations also consumed the budget. Eight single-agent workflows and 16 multi-agent workflows exhausted the 24-call calculator budget. Response-format errors caused three further single-agent failures and four multi-agent failures. The remaining single-agent failure involved a missing matching calculation. These results measure calculation-interface adherence alongside document interpretation and evidence control.

One single-agent issuance attempt selected the feasible, currently approved R90 alternative but attached an unsupported source link to the fixed sale proceeds. Its premise set also lacked a matching successful calculation. The adapter blocked the attempt; scoring retained it as an inadmissible raw proposal. The enforced issuance invariant recorded zero violations for all workflows. The deterministic, single-agent and multi-agent workflows completed six, four and three of the six withdrawal trajectories, respectively. The other five model attempts failed before initial approval.

Six response-format failures lacked complete final usage totals, two in the single-agent workflow and four in the multi-agent workflow, affecting five case repetitions. Reported costs cover 37 matched records per workflow, while decision analysis includes all 42. The multi-agent cost was 1.72 times the single-agent cost within that subset (\cref{tab:narrative-live}), including metered failures and required holds.

The revised narrative interface completed all 126 assignments, with 42 correct decisions and all six withdrawal trajectories per workflow (\cref{tab:narrative-interface}). This revision changed extraction and calculator interaction under the retained case policies; the Pareto and answer-partition rules in \cref{alg:decision} were assessed separately in deterministic constructions. Failures, validation errors and enforced gate violations were zero. All model proposals were admissible at their evidence versions. Each workflow asked 30 relevant questions and zero redundant questions. Every calculator event returned a single feasible candidate, so these results concern extraction and evidence control.

\input{3-assets/1-User/tables/narrative-interface-results}

The revised comparison has complete usage accounting for all 42 records per workflow. The multi-agent workflow incurred 2.64 times the single agent's reported model cost, including the costs of required holds (\cref{tab:narrative-interface}). Both used the shared arithmetic-receipt interface. The live run required zero correction attempts.

Independent reconstruction checked all 252 narrative records and their event chronology. The initial run contained 24 deterministic, 14 single-agent and seven multi-agent issued outputs; the revised run contained 24 per workflow. Across these 117 outputs, the checker found zero violations of the supplied financial, evidence, provenance and approval requirements. These results assess the retained emissions under the authored requirements; domain validity remains with the case-policy review.

\Cref{fig:workflow-results} contrasts the initial conformance gaps with the revised 42/42 reference outcomes for every workflow. Revised model expenditures across all 42 assignments are USD~0, USD~2.13 and USD~5.63 for deterministic, single-agent and multi-agent execution. On these two observed dimensions, the deterministic workflow strictly dominates both model arrangements and the single agent strictly dominates the multi-agent arrangement: conformance is equal and model expenditure is lower. This descriptive comparison includes correct recommendations and required holds and retains the experimental grammar, supplied policy and resource-accounting boundary.

\input{3-assets/1-User/figures/workflow-results}

\section{Discussion}\label{sec:discussion}

The simulations draw attention to choices made before an agent begins to optimise. In the constructed cases, an omitted liability led to a liquidity shortfall, omitted manager terms breached supervision capacity, and a mistranslated weight changed the selected service. Duplicating a client facet also changed the objective's behaviour. These examples suggest a practical reason to review disclosure and mandate translation alongside execution. Proposition~\ref{prop:indistinguishability} gives the corresponding information limit: selecting a client-optimal outcome across states with disjoint optima requires complete decision inputs that distinguish those states.

Nested delegation offers a way to organise that review around the manager's dual role. The manager agrees the mandate with the client and translates it into instructions for the system. In the proposed framework, mandate review addresses the objectives, scales, service terms and evidence criteria; execution review checks their application and current authorization. Making those responsibilities explicit could help reviewers trace a questionable recommendation to the choices or information on which it depends.

In the joint objective, benchmark floors preserve both declared evaluations relative to a reference service, and positive weights select a Pareto-efficient pair among the alternatives satisfying the constraints and floors. Law and mandate determine those alternatives before scalarisation. This ordering gives a place in the specification to MiFID~II Article~24's client-interest duties and Article~54(1)'s retained responsibility for automated suitability assessment \parencites[paras.~1 and 10]{esma2026mifid24}[art.~54(1)]{europeancommission2017suitability}. The choice of utility scales, benchmarks and commercial criteria still calls for an assessment against those duties. Even the interpretation of $\lambda$ depends on the scales: rescaling manager value while holding the weight fixed changes the effective trade-off.

The manager-floor example makes the stakes of choosing a reference service concrete. Preserving manager value at $1/2$ selects $(C,\omega_1)$ with client value $3/4$, while excluding $(B,\omega_2)$ with client value $1$ and manager value $1/4$. The weight concession is zero; the benchmark and total concessions are $1/4$. Under the proposed review rule in Equation~\eqref{eq:authorization}, that concession calls for policy review and supporting client-benefit evidence. A necessary delivery restriction instead enters admissibility and changes the comparator. Client-confirmed preferences, agreed service terms and domain assessment provide grounds for accepting the policy or revising the reference service.

The economic simulation illustrates a similar tension using constructed fees, personnel costs, supervision hours and service-quality values. In six of eight states, the declared weight selected a higher service tier than the client-best pair, including with a client weight close to one. The floors were nonbinding in those cases, so the reported client amounts equal the total concessions. In the 2022 simulation, expected benchmark preservation also coexisted with a lower subsequent client outcome and higher manager net income. That euro comparison applies the same household-payment adjustment and service valuation to both services; it measures a different quantity from the risk-adjusted decision criterion. The example suggests that review should extend to the forecasts and valuations supporting the mandate.

Service classification matters as well. In the Swiss illustration, the common hold policy adds an institutional restriction to the mapped FinSA information duty; the EU necessary-information gate follows Article~54(8) \parencites[arts.~11--14]{switzerland2018finsa}[art.~54(8)]{europeancommission2017suitability}. A candidate can remain financially feasible while a change in the required-information classification changes permission to proceed. Recording the basis of each restriction could make such decisions easier to explain. The same applies to the distinction between evidence sufficiency and permission to collect it: the specification records both the required information and its permitted acquisition routes \parencites{esma1762missinginfo}[arts.~5--6]{europeanunion2016gdpr}[arts.~6--7]{switzerland2020fadp}.

Clarification raises further questions about how the mandate handles uncertainty. Statewise Pareto filtering removes the demonstrated dominated alternatives, including the zero-product case. Candidate-specific assessment allows an eligible maximiser to proceed while another awaits clarification. The answer-partition test detects the supplied partial-answer consequence, and the client-criterion diagnostic detects a gain from changing candidates even at equal euro losses. These properties hold for the declared states, facets, answer partitions and thresholds. The partial-answer example is particularly relevant to mandate design: the componentwise robust comparison can reverse as evidence arrives, even when every fully resolved state initially selects the same candidate. Whether that behaviour suits the client depends on the chosen treatment of uncertainty.

There are several reasons to seek further information. Required gaps govern permission to proceed; optional answers may improve selection, reveal dominance or restore feasibility. The restricted selector evaluates declared conditional gains under a fixed policy. An answer that would restore an empty feasible set still needs human resolution, while a discretionary question about the full joint objective would need assessment against both parties' evaluations and the workflow. The simulation assigns costs and source owners to questions. Choosing a sequence of questions would require a further account of their costs and possible answers, with probabilities for an expected-value treatment. Calibration on representative mandates could examine consequential omissions, question burden, unnecessary holds and review effort before evaluating a chosen policy on held-out cases.

After a recommendation, evidence changes may call for a fresh assessment. The withdrawal and replacement checks distinguish lost support from surviving independent evidence; the currency example shows how a changed valuation can leave an unchanged obligation underfunded. Such checks could contribute to the testing, monitoring and allocation of responsibility addressed by FINMA \parencite[secs.~2.1 and 2.4]{finma2024ai}. They also give engineers concrete events that should trigger reassessment or renewed approval.

The narrative runs illustrate how implementation choices can affect execution. Rejected inputs and repeated calculations exhausted budgets in the initial interface, and the adapter blocked a proposal carrying unsupported attribution and missing calculation support. Typed extraction, clearer instructions and shared arithmetic receipts subsequently reached all reference outcomes on the same development cases. This experience suggests that the interface deserves attention alongside the decision rule. Because those changes were introduced together and the correction mechanism remained unused, their individual contributions remain open.

In the instruction comparison, professional prose, the explicit joint objective and nested-delegation instructions reached the same prescribed decisions; professional prose also matched nested delegation on question count. All three received the declared trade-off, both floors, source identities and common enforcement. Their agreement illustrates several ways of communicating the same specification. The representation interventions changed its content and produced departures from the original mandate. Recovering missing inputs used a supplied field schema and authoritative source routes. Recognising an omitted field, choosing an appropriate scale or justifying a reference service would call on professional judgment at an earlier stage.

Workflow design raises a related question about the value of additional roles. In the revised narrative comparison, deterministic execution and the single agent achieved the same conformance as the multi-agent arrangement at lower measured model cost. The two facet-labelled calls and coordinator each received the complete policy and common numerical support, and every revised calculator event had one feasible candidate. The observed agreement therefore concerns extraction and evidence-aware sequencing; the analytical constructions examine selection among competing feasible alternatives. Giving specialists complementary information, with budget-matched repeated single-agent checking as a comparison, could help assess when decomposition adds value.

The service workflow $\omega$ and the agent architecture would both need consideration in an applied system. The simulated service tiers differ in fees, quality and supervisory hours, while the instruction treatments hold architecture fixed. The narrative comparison varies architecture and records model expenditure. Relating these choices would require measurements of professional effort, infrastructure cost and delivered service quality. Additional roles may be worthwhile when their information or expertise improves the service enough to justify those costs. The proposed outcome functions offer a place to express and examine that judgment.

\section{Conclusion}\label{sec:conclusion}

This paper formulates the task of a delegated AI system as joint maximisation under binding constraints within a nested client--manager--AI relationship. The formulation represents client and manager interests through separate outcome functions, with explicit constraints, reference floors and trade-off weights. It connects responsibility for the mandate to its translation into an executable decision specification. Concession accounting makes the client consequences of the floors and weights available for review, while the distinction between disclosure, translation and execution helps locate possible departures from the intended mandate.

The analytical constructions and simulation illustrate how this framework can guide calculations, clarification questions and approval controls. Within the simulated settings, changes in represented information, constraints and policy choices affected the available or selected decisions. These observations suggest that making such choices explicit could support more structured mandate design and review. The underlying proposal is independent of a particular model or agent architecture and may provide a useful basis for comparable implementations. Professional specification studies and field evaluations could assess whether, and under which conditions, this approach improves decision quality, oversight and client outcomes.

\section{Limitations and Future Research}\label{sec:limitations-future}

The simulation instantiates the joint objective with constructed mandates, preferences, service terms and growth scenarios. Public exchange-rate and yield observations anchor selected market inputs; they supply current retrieval vintages of historical observations. Client evaluation scales, reference service, quality values and managerial commercial interests require domain validation. The source-linked legal analysis supplies the context for institution-specific mandate and processing review. Strategic disclosure, hidden effort and incentive-compatible implementation require behavioural models and observations of how the parties respond to the specification.

Information degradation depends on the disclosure and specification maps and available supplementary evidence. The indistinguishability result assumes identical complete decision inputs and fixed policy primitives. Finite state enumeration, omitted-state discovery and evidence sufficiency bound the procedure's applicability. Statewise Pareto filtering and the client-criterion trigger depend on declared cardinal utilities. Flat Nash products can conceal an answer-dependent dominance change even at zero clarification thresholds; undominated issuance concerns the current support. Optional questions that restore an empty feasible set in the restricted client selector require human evidence resolution. The clarification policy examines one-input partitions and possible gains. Joint inputs can conceal a selection gain even when every individual consequence is zero; question probabilities, costs and compound partitions remain open. Applying the legal mapping to an actual mandate requires current provider authorization or exemption, client category, territorial facts and processing arrangements, alongside production identity and storage controls.

The structured pilot spans six families sharing three motivating source clusters; the narratives span two source-linked families with controlled grammar. Repetitions measure within-case consistency. All workflows receive authored policies and numerical support. The revised narrative interface was developed from failures on the same cases used for its evaluation. All 114 revised calculator events returned one feasible candidate, comprising 57 R90 and 57 R180 results. The deterministic parser is tailored to the document grammar. Initial model costs cover 37 matched records per workflow because six failures lack final usage totals; revised costs cover all 42. The deterministic policy checks establish properties of the Pareto and partition procedure, while the historical model comparisons retain their original policy scope.

The joint-objective comparison has four base mandates, eight dependent conditions per mandate and one repetition per instruction treatment. Evaluation mandates were reserved from prompt development and share their four templates and weights with the development mandates; an interrupted preliminary execution exposed a numerical grading defect, corrected by standardising monetary inputs to cents before the reported comparison. The final treatment instructions were retained. Source records present explicit numerical fields, source owners and bounded question routes, and the common approval layer rejects unsupported or inadmissible proposals. The observed parity concerns that shared environment. Broader language variation, independently authored mandates and budget-matched controls would assess transfer; related conditions should remain grouped in allocation and uncertainty estimates.

A prospective reviewer comparison would assess the framework against a competent governance checklist covering the same controls. Reviewers would diagnose matched defects in mandate representation, requirements translation, premise support and current approval. Decision accuracy, responsibility assignment, unnecessary holds and review effort would remain separate outcomes. Practitioner responses and independent domain assessment of case labels would establish whether organizing the controls through nested delegation improves professional specification and review. Field evaluation would examine client outcomes after oversight and implementation costs.

A prospective screening extension would specify disclosure menus and test truth-telling and participation conditions using client utilities, disclosure costs, outside options and outcomes of possible deviations \parencite{rothschildstiglitz1976,mirrlees1976incentives}. Controlled behavioural studies would examine \emph{generative moral hazard}, the hypothesis that omitted checks can remain concealed behind a plausible recommendation, and its dependence on effort and generation incentives \parencite{holmstrom1979moral,kalai2025hallucinate}.

\printbibliography

\swDeclarationsPage
\end{document}

%% file: 1-system/0-meta.tex
\newcommand{\papertitle}{Multi-Agent AI as a Nested Principal-Agent Problem in Private Wealth Management}
\newcommand{\papersubtitle}{Mandate Representation and Evidence Control in Switzerland, Germany and Austria}

\newcommand{\papercitationauthors}{Kurz, W. and Magg, R. and Kollberg, F. and Stricker, W. and Marx, S. and Reinhardt, F. and Dedi\'{c}, V.}

\newcommand{\paperauthors}{%
  Walter Kurz\textsuperscript{1}, %
  Reinhard Magg\textsuperscript{1}, %
  Florian Kollberg\textsuperscript{2},
  Wojtek Stricker\textsuperscript{1}, %
  Stefan Marx\textsuperscript{2}, %
  Frank Reinhardt\textsuperscript{2}, %
  Velimir Dedi\'{c}\textsuperscript{3}%
}

\newcommand{\paperaffiliations}{%
  \textsuperscript{1}Swissi Institute for AI,
  \href{mailto:office@swissi-ai.institute}{office@swissi-ai.institute};
  \textsuperscript{2}Hochschule f{\"u}r Wirtschaft und Umwelt N{\"u}rtingen-Geislingen\\
  \textsuperscript{3}Faculty of Information Technology and Engineering (FITI), Belgrade%
}

\newcommand{\paperdeclarationaffiliations}{\begin{tabular}[t]{@{}p{\linewidth}@{}}
  \resizebox{\linewidth}{!}{%
  \textsuperscript{1}Swissi Institute for AI,
  \href{mailto:office@swissi-ai.institute}{office@swissi-ai.institute};
  \textsuperscript{2}Hochschule f{\"u}r Wirtschaft und Umwelt N{\"u}rtingen-Geislingen%
  }\\
  \textsuperscript{3}Faculty of Information Technology and Engineering (FITI), Belgrade%
\end{tabular}}

\newcommand{\papercorresponding}{Walter Kurz, \href{mailto:me@walterkurz.com}{me@walterkurz.com}}

\newcommand{\paperfunding}{This research received no external funding.}
\newcommand{\paperconflictsofinterest}{The authors declare no conflicts of interest.}
\newcommand{\paperdataavailability}{The authors retain the source code, constructed mandates, market inputs, recorded responses, experimental configurations and verification instructions. Access to these research records is restricted. Market inputs include source URLs and content hashes.}
\newcommand{\paperaitools}{GPT-6.1 Astra and Claude Opus 5.5 were used for the simulation.}
\newcommand{\paperauthorcontributions}{All authors contributed equally to this work.}

%% file: 3-assets/1-User/figures/nested-agency.tex
\begin{figure}[htbp]
\centering
\begin{tikzpicture}[
  box/.style={draw, text width=6.0cm, minimum height=1.15cm,
    align=center, font=\small, inner sep=6pt},
  conn/.style={->, >=stealth, semithick},
  edge/.style={font=\footnotesize, align=center, text width=3.4cm},
]
\node[box] (client) at (0,0)
  {\textbf{Client $P_1$}\\Client interests and private circumstances};
\node[box] (manager) at (0,-2.6)
  {\textbf{Wealth manager $A_1=P_2$}\\Manager interests\\Mandate and AI policy};
\node[box] (system) at (0,-5.2)
  {\textbf{AI system $A_2$}\\Received evidence and objectives\\Constrained recommendation};

\draw[conn] ([xshift=-1.8cm]client.south) -- ([xshift=-1.8cm]manager.north);
\node[edge] at (-3.7,-1.3) {Disclosed context\\and mandate};
\draw[conn] ([xshift=1.8cm]manager.north) -- ([xshift=1.8cm]client.south);
\node[edge] at (3.7,-1.3) {Documented\\advice};
\draw[conn] ([xshift=-1.8cm]manager.south) -- ([xshift=-1.8cm]system.north);
\node[edge] at (-3.7,-3.9) {Specified objectives,\\policy and budgets};
\draw[conn] ([xshift=1.8cm]system.north) -- ([xshift=1.8cm]manager.south);
\node[edge] at (3.7,-3.9) {Proposed advice,\\premises and escalations};

\draw[<->, >=stealth, semithick, dashed]
  (client.west) -- (-6.3,0) -- (-6.3,-5.2) -- (system.west);
\node[font=\footnotesize, rotate=90, fill=white, inner sep=4pt]
  at (-6.3,-2.6) {Client-context elicitation};
\end{tikzpicture}
\caption{Nested delegation and client-context elicitation.}
\label{fig:nested-agency}
{\raggedright\fignote{The wealth manager acts as the client's agent ($A_1$) and the AI system's principal ($P_2$). Solid arrows show the two delegation relationships; the dashed connection shows permitted client-context elicitation. Source: the authors' conceptual model.}}
\end{figure}
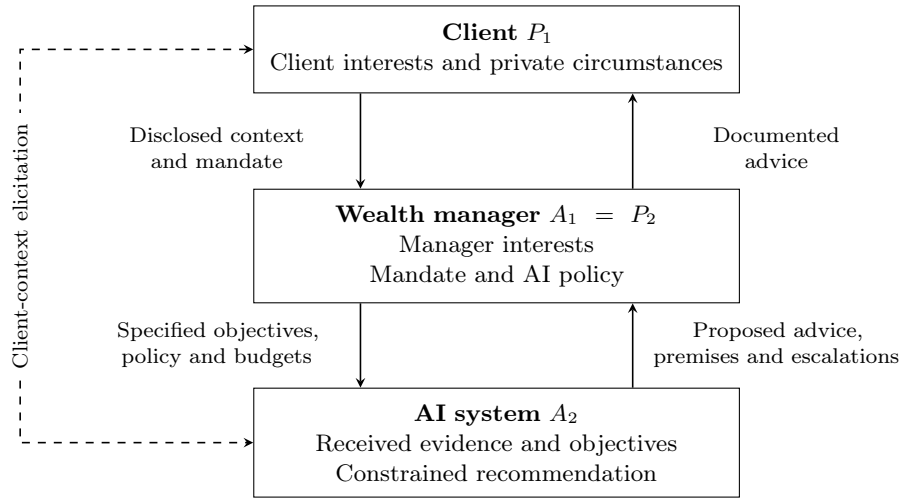

%% file: 3-assets/1-User/figures/recommendation-flow.tex
\begin{figure}[htbp]
\centering
\begin{tikzpicture}[
  box/.style={draw, text width=5.5cm, minimum height=0.95cm,
    align=center, font=\small, inner sep=5pt},
  side/.style={draw, text width=3.0cm, minimum height=0.95cm,
    align=center, font=\footnotesize, inner sep=5pt},
  conn/.style={->, >=stealth, semithick},
  edge/.style={font=\footnotesize, fill=white, inner sep=2pt},
]
\node[box] (evidence) at (0,0) {\textbf{Purpose-scoped evidence}};
\node[box] (gates) at (0,-1.8)
  {\textbf{Mandatory checks}\\Lawfulness, provenance and suitability};
\node[box] (context) at (0,-3.6)
  {\textbf{Joint-state calculation}\\Feasible sets and Pareto selection};
\node[box] (facets) at (0,-5.4)
  {\textbf{Answer-partition consequences}\\Candidate-specific eligibility};
\node[box] (nash) at (0,-7.2)
  {\textbf{Review-eligible maximisers}\\Current evidence and policy version};
\node[box] (human) at (0,-9.0)
  {\textbf{Human review}\\Approve or hold};
\node[box] (advice) at (0,-10.8) {\textbf{Documented recommendation}};

\draw[conn] (evidence.south) -- (gates.north);
\draw[conn] (gates.south) -- node[edge,right] {pass} (context.north);
\draw[conn] (context.south) -- node[edge,right] {non-empty set} (facets.north);
\draw[conn] (facets.south) -- node[edge,right] {$\mathcal R(E)\ne\varnothing$} (nash.north);
\draw[conn] (nash.south) -- (human.north);
\draw[conn] (human.south) -- node[edge,right] {approved} (advice.north);

\node[side] (gatehold) at (-5.7,-1.8)
  {\textbf{Hold and resolve}\\Clarify or revise;\\refer as needed};
\draw[conn] (gates.west) -- node[edge,above] {fail} (gatehold.east);
\draw[conn] (gatehold.north) -- (-5.7,0)
  -- node[edge,above,align=center] {Revised evidence\\or scope} (evidence.west);
\draw[conn] (context.west) -- node[edge,below] {empty set}
  (-5.7,-3.6) -- (gatehold.south);

\node[side] (question) at (6.7,-5.4)
  {\textbf{Lawful question}\\or human clarification};
\draw[conn] (facets.east) -- node[edge,above] {$\mathcal R(E)=\varnothing$} (question.west);
\draw[conn] (question.north) -- (6.7,0)
  -- node[edge,above,align=center] {Updated\\evidence} (evidence.east);

\node[side] (contexthold) at (6.7,-9.0)
  {\textbf{Hold and refer}\\Decision context\\unresolved};
\draw[conn] (question.south) --
  node[edge,align=center] {Refusal, timeout\\or insufficient answer}
  (contexthold.north);
\end{tikzpicture}
\caption{Evidence-dependent calculation and version-bound approval in the restricted client selector.}
\label{fig:recommendation-flow}
{\raggedright\fignote{$\mathcal R(E)$ is the set of review-eligible maximisers at evidence state $E$. Approval binds a candidate to the current evidence and policy version. Source: the authors' recommendation procedure.}}
\end{figure}
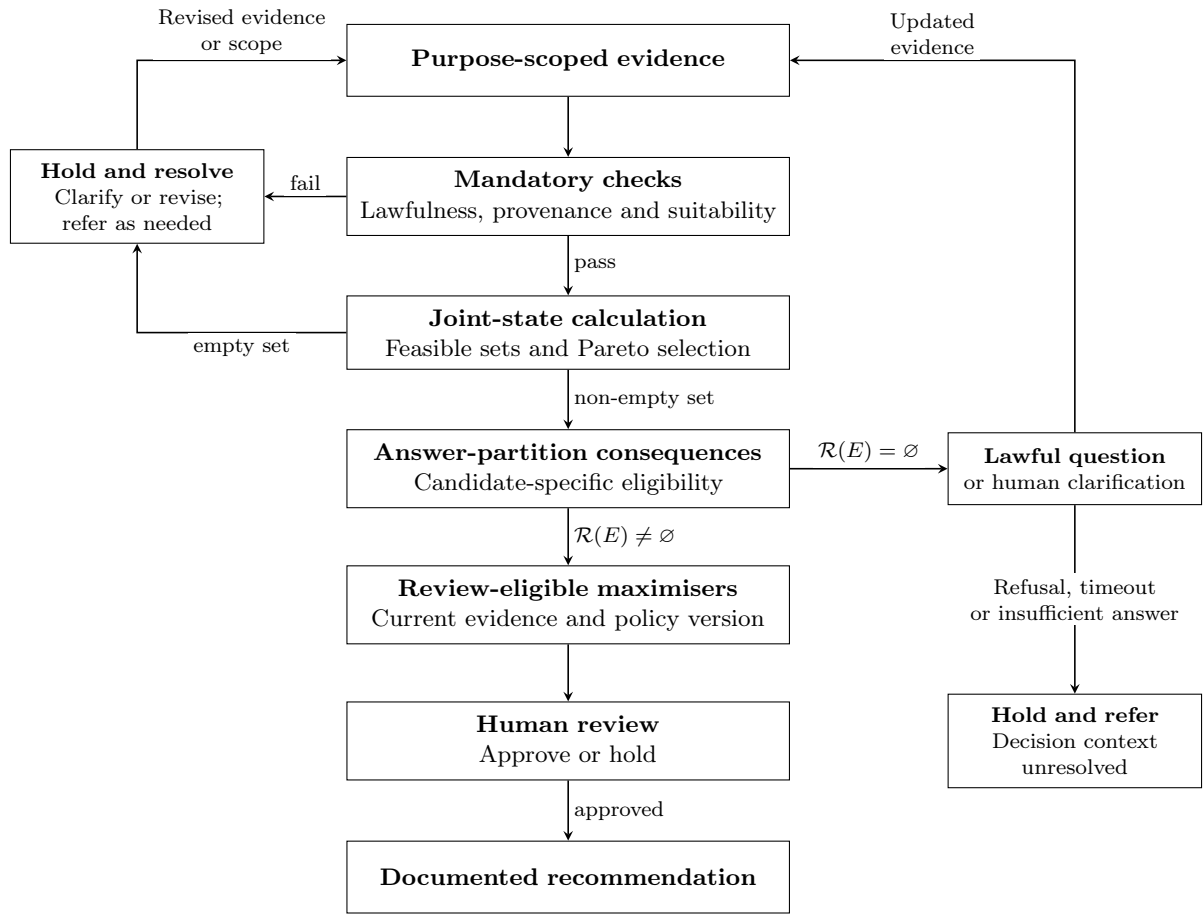

%% file: 3-assets/1-User/figures/joint-tradeoff.tex
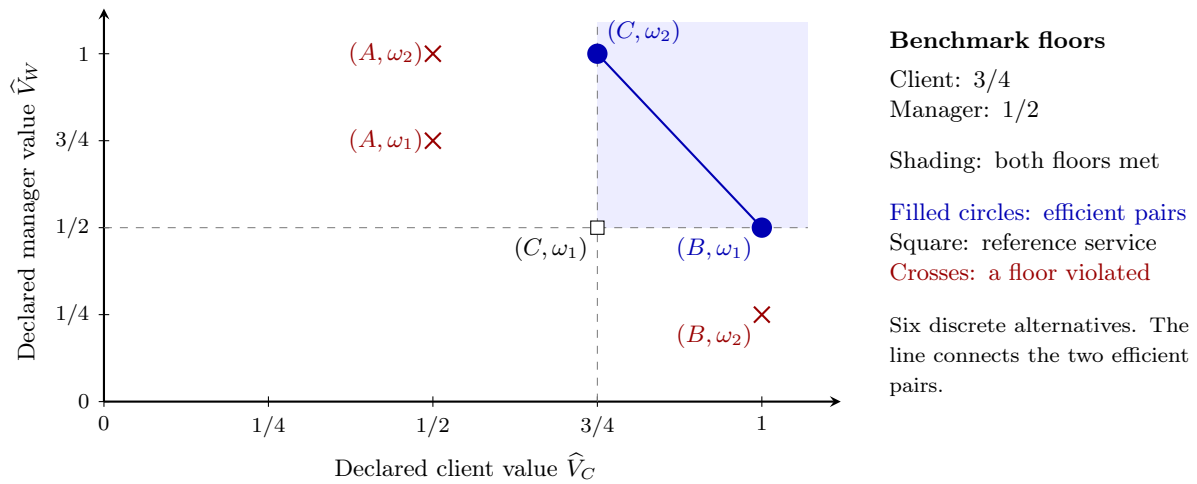
\begin{figure}[htbp]
\centering
\input{3-assets/1-User/figures/plots/joint-tradeoff}
\caption{Client and manager values of the constructed alternatives.}
\label{fig:joint-tradeoff}
{\raggedright\fignote{The six portfolio--workflow pairs use the client values and workflow burdens defined in the text. Shading marks the benchmark region; the connecting segment indicates the trade-off between the two efficient discrete alternatives. Source: exact joint-objective construction.}}
\end{figure}

%% file: 3-assets/1-User/figures/plots/joint-tradeoff.tex
\input{3-assets/1-User/figures/plots/result-values.tex}
\begin{tikzpicture}[x=8.7cm,y=4.6cm,font=\small,>=stealth]
  \fill[blue!7] (\JClientCone,\JManagerCone) rectangle (1.07,1.09);
  \draw[->,thick] (0,0) -- (1.12,0);
  \draw[->,thick] (0,0) -- (0,1.13);
  \foreach \v/\lab in {0/0,0.25/{1/4},0.5/{1/2},0.75/{3/4},1/1} {
    \draw (\v,0.015) -- (\v,-0.015) node[below,font=\footnotesize] {$\lab$};
    \draw (0.008,\v) -- (-0.008,\v) node[left,font=\footnotesize] {$\lab$};
  }
  \node[below] at (0.55,-0.12) {Declared client value $\widehat V_C$};
  \node[rotate=90] at (-0.12,0.55) {Declared manager value $\widehat V_W$};
  \draw[dashed,gray] (\JClientCone,0) -- (\JClientCone,1.09);
  \draw[dashed,gray] (0,\JManagerCone) -- (1.07,\JManagerCone);
  \draw[thick,blue!70!black] (\JClientCtwo,\JManagerCtwo) -- (\JClientBone,\JManagerBone);
  \node[circle,fill=blue!70!black,inner sep=2.7pt] at (\JClientCtwo,\JManagerCtwo) {};
  \node[above right,text=blue!70!black] at (\JClientCtwo,\JManagerCtwo) {$(C,\omega_2)$};
  \node[circle,fill=blue!70!black,inner sep=2.7pt] at (\JClientBone,\JManagerBone) {};
  \node[below left,text=blue!70!black] at (\JClientBone,\JManagerBone) {$(B,\omega_1)$};
  \node[rectangle,draw,fill=white,inner sep=2.5pt] at (\JClientCone,\JManagerCone) {};
  \node[below left] at (\JClientCone,\JManagerCone) {$(C,\omega_1)$};
  \foreach \x/\y in {\JClientAone/\JManagerAone,\JClientAtwo/\JManagerAtwo,\JClientBtwo/\JManagerBtwo} {
    \draw[thick,red!60!black] (\x-0.012,\y-0.023) -- (\x+0.012,\y+0.023);
    \draw[thick,red!60!black] (\x-0.012,\y+0.023) -- (\x+0.012,\y-0.023);
  }
  \node[left,text=red!60!black] at (\JClientAone,\JManagerAone) {$(A,\omega_1)$};
  \node[left,text=red!60!black] at (\JClientAtwo,\JManagerAtwo) {$(A,\omega_2)$};
  \node[below left,text=red!60!black] at (\JClientBtwo,\JManagerBtwo) {$(B,\omega_2)$};
  \node[anchor=north west,align=left,text width=4.0cm] at (1.18,1.09)
    {\textbf{Benchmark floors}\\[3pt]
     Client: $3/4$\\Manager: $1/2$\\[8pt]
     Shading: both floors met\\[8pt]
     \textcolor{blue!70!black}{Filled circles: efficient pairs}\\
     Square: reference service\\
     \textcolor{red!60!black}{Crosses: a floor violated}\\[8pt]
     {\footnotesize Six discrete alternatives. The line connects the two efficient pairs.}};
\end{tikzpicture}

%% file: 3-assets/1-User/figures/weight-sensitivity.tex
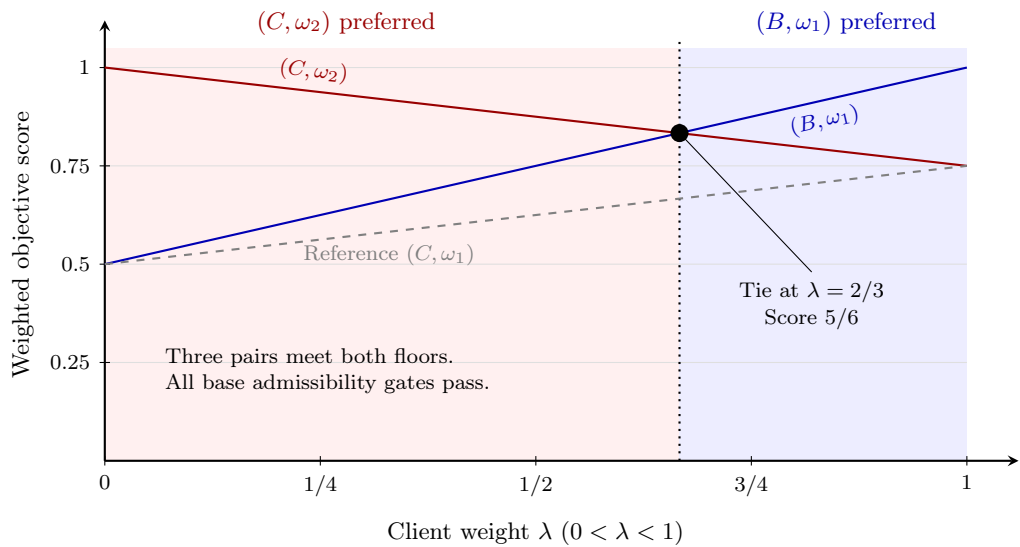
\begin{figure}[htbp]
\centering
\input{3-assets/1-User/figures/plots/weight-sensitivity}
\caption{Objective scores across client weights.}
\label{fig:weight-sensitivity}
{\raggedright\fignote{The three pairs meet both benchmark floors, with all base admissibility gates satisfied. Endpoint values show limits of the specified open weight interval. Source: exact joint-objective construction.}}
\end{figure}

%% file: 3-assets/1-User/figures/plots/weight-sensitivity.tex
\input{3-assets/1-User/figures/plots/result-values.tex}
\begin{tikzpicture}[x=11.4cm,y=5.2cm,font=\small,>=stealth]
  \fill[red!6] (0,0) rectangle ({2/3},1.05);
  \fill[blue!7] ({2/3},0) rectangle (1,1.05);
  \foreach \v in {0.25,0.5,0.75,1} {
    \draw[gray!25] (0,\v) -- (1,\v);
    \draw (0.006,\v) -- (-0.006,\v) node[left,font=\footnotesize] {\v};
  }
  \draw[->,thick] (0,0) -- (1.06,0);
  \draw[->,thick] (0,0) -- (0,1.12);
  \foreach \v/\lab in {0/0,0.25/{1/4},0.5/{1/2},0.75/{3/4},1/1} {
    \draw (\v,0.012) -- (\v,-0.012) node[below,font=\footnotesize] {$\lab$};
  }
  \node[rotate=90] at (-0.095,0.55) {Weighted objective score};
  \node[below] at (0.5,-0.13) {Client weight $\lambda$ $(0<\lambda<1)$};
  \draw[dotted,thick] ({2/3},0) -- ({2/3},1.07);
  \draw[thick,red!60!black] (0,\JManagerCtwo) -- (1,\JClientCtwo);
  \draw[thick,blue!70!black] (0,\JManagerBone) -- (1,\JClientBone);
  \draw[thick,dashed,gray] (0,\JManagerCone) -- (1,\JClientCone);
  \node[above,text=red!60!black] at (0.28,1.06) {$(C,\omega_2)$ preferred};
  \node[above,text=blue!70!black] at (0.86,1.06) {$(B,\omega_1)$ preferred};
  \node[circle,fill=black,inner sep=2.4pt] at ({2/3},{5/6}) {};
  \draw[thin] ({2/3},{5/6}) -- (0.82,0.48);
  \node[anchor=north,align=center,font=\footnotesize] at (0.82,0.48)
    {Tie at $\lambda=2/3$\\Score $5/6$};
  \node[above,rotate=-6,text=red!60!black,font=\footnotesize] at (0.24,0.94) {$(C,\omega_2)$};
  \node[below,rotate=13,text=blue!70!black,font=\footnotesize] at (0.83,0.915) {$(B,\omega_1)$};
  \node[below,text=gray,font=\footnotesize] at (0.33,0.575) {Reference $(C,\omega_1)$};
  \node[anchor=north west,align=left,font=\footnotesize,text width=4.5cm] at (0.06,0.31)
    {Three pairs meet both floors.\\All base admissibility gates pass.};
\end{tikzpicture}

%% file: 3-assets/1-User/figures/nested-tradeoff.tex
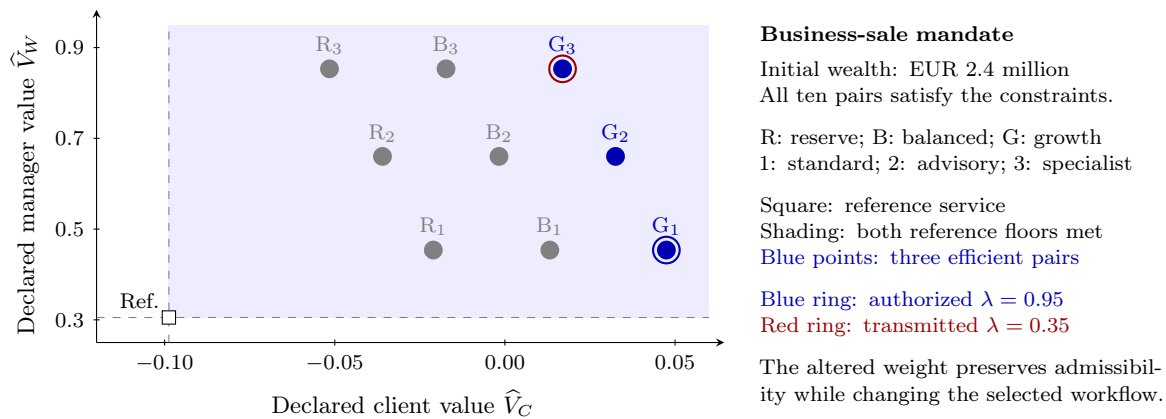
\begin{figure}[htbp]
\centering
\input{3-assets/1-User/figures/plots/nested-tradeoff}
\caption{Economic trade-offs and the effect of a mistranslated joint weight.}
\label{fig:nested-tradeoff}
{\raggedright\fignote{Complete heldout business-sale case, December 2021 decision. Points are the nine portfolio--workflow pairs and the reference service; both axes use the declared outcome scales. Client value includes the forecast risk penalty and service value. The two rings identify the choices under the authorized and altered weights. Each point is a discrete alternative. Source: frozen simulation inputs and exact valuation.}}
\end{figure}

%% file: 3-assets/1-User/figures/plots/nested-tradeoff.tex
\begin{tikzpicture}[font=\small,>=stealth]
  \fill[blue!7] (0.95662,0.33100) rectangle (8.1,4.2);
  \draw[->] (0,0) -- (8.3,0);
  \draw[->] (0,0) -- (0,4.4);
  \draw[dashed,gray] (0.95662,0) -- (0.95662,4.2);
  \draw[dashed,gray] (0,0.33100) -- (8.1,0.33100);
  \draw (0.90000,0.04) -- (0.90000,-0.04) node[below,font=\footnotesize] {$ -0.10 $};
  \draw (3.15000,0.04) -- (3.15000,-0.04) node[below,font=\footnotesize] {$ -0.05 $};
  \draw (5.40000,0.04) -- (5.40000,-0.04) node[below,font=\footnotesize] {$ 0.00 $};
  \draw (7.65000,0.04) -- (7.65000,-0.04) node[below,font=\footnotesize] {$ 0.05 $};
  \draw (0.04,0.30000) -- (-0.04,0.30000) node[left,font=\footnotesize] {$ 0.3 $};
  \draw (0.04,1.50000) -- (-0.04,1.50000) node[left,font=\footnotesize] {$ 0.5 $};
  \draw (0.04,2.70000) -- (-0.04,2.70000) node[left,font=\footnotesize] {$ 0.7 $};
  \draw (0.04,3.90000) -- (-0.04,3.90000) node[left,font=\footnotesize] {$ 0.9 $};
  \node[below] at (4.05,-0.5) {Declared client value $\widehat V_C$};
  \node[rotate=90] at (-0.95,2.1) {Declared manager value $\widehat V_W$};
  \node[circle,fill=gray,inner sep=2.5pt] at (5.32181,2.46100) {};
  \node[above=2pt,text=gray,font=\footnotesize] at (5.32181,2.46100) {$\mathrm{B}_2$};
  \node[circle,fill=gray,inner sep=2.5pt] at (4.62126,3.61900) {};
  \node[above=2pt,text=gray,font=\footnotesize] at (4.62126,3.61900) {$\mathrm{B}_3$};
  \node[circle,fill=gray,inner sep=2.5pt] at (5.99424,1.22260) {};
  \node[above=2pt,text=gray,font=\footnotesize] at (5.99424,1.22260) {$\mathrm{B}_1$};
  \node[circle,fill=blue!70!black,inner sep=2.5pt] at (6.86274,2.46100) {};
  \node[above=2pt,text=blue!70!black,font=\footnotesize] at (6.86274,2.46100) {$\mathrm{G}_2$};
  \node[circle,fill=blue!70!black,inner sep=2.5pt] at (6.16219,3.61900) {};
  \node[above=2pt,text=blue!70!black,font=\footnotesize] at (6.16219,3.61900) {$\mathrm{G}_3$};
  \draw[red!60!black,thick] (6.16219,3.61900) circle[radius=5pt];
  \node[circle,fill=blue!70!black,inner sep=2.5pt] at (7.53517,1.22260) {};
  \node[above=2pt,text=blue!70!black,font=\footnotesize] at (7.53517,1.22260) {$\mathrm{G}_1$};
  \draw[blue!70!black,thick] (7.53517,1.22260) circle[radius=5pt];
  \node[rectangle,draw,fill=white,inner sep=2.5pt] at (0.95662,0.33100) {};
  \node[above left,font=\footnotesize] at (0.95662,0.33100) {Ref.};
  \node[circle,fill=gray,inner sep=2.5pt] at (3.78088,2.46100) {};
  \node[above=2pt,text=gray,font=\footnotesize] at (3.78088,2.46100) {$\mathrm{R}_2$};
  \node[circle,fill=gray,inner sep=2.5pt] at (3.08033,3.61900) {};
  \node[above=2pt,text=gray,font=\footnotesize] at (3.08033,3.61900) {$\mathrm{R}_3$};
  \node[circle,fill=gray,inner sep=2.5pt] at (4.45330,1.22260) {};
  \node[above=2pt,text=gray,font=\footnotesize] at (4.45330,1.22260) {$\mathrm{R}_1$};
  \node[anchor=north west,align=left,text width=5.4cm,font=\footnotesize] at (8.65,4.3)
    {\textbf{Business-sale mandate}\\[3pt]
     Initial wealth: EUR 2.4 million\\
     All ten pairs satisfy the constraints.\\[6pt]
     R: reserve; B: balanced; G: growth\\
     1: standard; 2: advisory; 3: specialist\\[6pt]
     Square: reference service\\
     Shading: both reference floors met\\
     \textcolor{blue!70!black}{Blue points: three efficient pairs}\\[6pt]
     \textcolor{blue!70!black}{Blue ring: authorized $\lambda=0.95$}\\
     \textcolor{red!60!black}{Red ring: transmitted $\lambda=0.35$}\\[6pt]
     The altered weight preserves admissibility while changing the selected workflow.};
\end{tikzpicture}

%% file: 3-assets/1-User/figures/nested-specification.tex
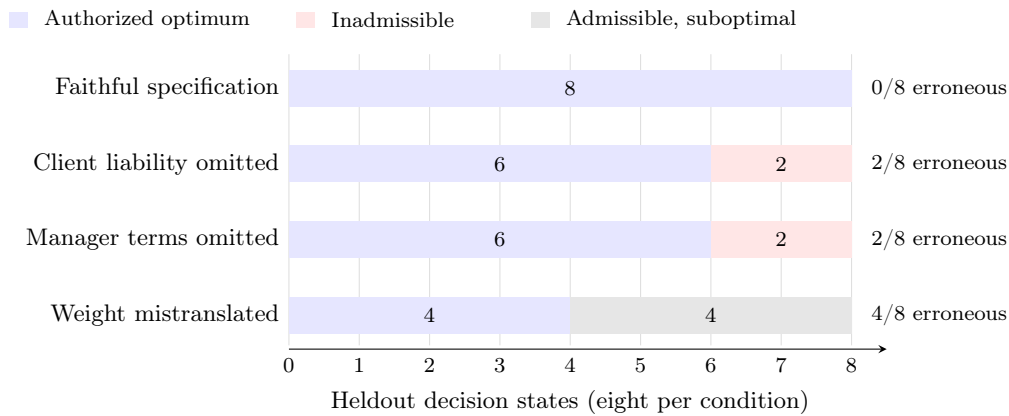
\begin{figure}[htbp]
\centering
\input{3-assets/1-User/figures/plots/nested-specification}
\caption{Consequences of controlled defects in the transmitted specification.}
\label{fig:nested-specification}
{\raggedright\fignote{Four mandate families, each evaluated in its complete and revised state. Exact optimization acts on each transmitted specification; outcomes are scored against the authoritative mandate. Categories partition the eight states in each row. Client omission substitutes zero liability and produces liquidity violations; manager omission substitutes zero hourly cost and 100 supervisory hours and produces capacity violations. Complete and revised states share their base mandate. Source: frozen heldout representation controls.}}
\end{figure}

%% file: 3-assets/1-User/figures/plots/nested-specification.tex
\begin{tikzpicture}[font=\small,>=stealth]
  \begin{scope}[x=0.93cm]
    \draw[->] (0,0.2) -- (8.5,0.2);
    \draw[gray!25] (0,0.25) -- (0,4.1);
    \node[below,font=\footnotesize] at (0,0.2) {0};
    \draw[gray!25] (1,0.25) -- (1,4.1);
    \node[below,font=\footnotesize] at (1,0.2) {1};
    \draw[gray!25] (2,0.25) -- (2,4.1);
    \node[below,font=\footnotesize] at (2,0.2) {2};
    \draw[gray!25] (3,0.25) -- (3,4.1);
    \node[below,font=\footnotesize] at (3,0.2) {3};
    \draw[gray!25] (4,0.25) -- (4,4.1);
    \node[below,font=\footnotesize] at (4,0.2) {4};
    \draw[gray!25] (5,0.25) -- (5,4.1);
    \node[below,font=\footnotesize] at (5,0.2) {5};
    \draw[gray!25] (6,0.25) -- (6,4.1);
    \node[below,font=\footnotesize] at (6,0.2) {6};
    \draw[gray!25] (7,0.25) -- (7,4.1);
    \node[below,font=\footnotesize] at (7,0.2) {7};
    \draw[gray!25] (8,0.25) -- (8,4.1);
    \node[below,font=\footnotesize] at (8,0.2) {8};
    \node[left,align=right] at (0,3.65) {Faithful specification};
    \fill[fill=blue!10] (0,3.41) rectangle (8,3.89);
    \node[text=black,font=\footnotesize] at (4.0,3.65) {8};
    \node[right,font=\footnotesize] at (8.15,3.65) {0/8 erroneous};
    \node[left,align=right] at (0,2.65) {Client liability omitted};
    \fill[fill=blue!10] (0,2.41) rectangle (6,2.89);
    \node[text=black,font=\footnotesize] at (3.0,2.65) {6};
    \fill[fill=red!10] (6,2.41) rectangle (8,2.89);
    \node[text=black,font=\footnotesize] at (7.0,2.65) {2};
    \node[right,font=\footnotesize] at (8.15,2.65) {2/8 erroneous};
    \node[left,align=right] at (0,1.65) {Manager terms omitted};
    \fill[fill=blue!10] (0,1.41) rectangle (6,1.89);
    \node[text=black,font=\footnotesize] at (3.0,1.65) {6};
    \fill[fill=red!10] (6,1.41) rectangle (8,1.89);
    \node[text=black,font=\footnotesize] at (7.0,1.65) {2};
    \node[right,font=\footnotesize] at (8.15,1.65) {2/8 erroneous};
    \node[left,align=right] at (0,0.65) {Weight mistranslated};
    \fill[fill=blue!10] (0,0.41) rectangle (4,0.89);
    \node[text=black,font=\footnotesize] at (2.0,0.65) {4};
    \fill[fill=gray!20] (4,0.41) rectangle (8,0.89);
    \node[text=black,font=\footnotesize] at (6.0,0.65) {4};
    \node[right,font=\footnotesize] at (8.15,0.65) {4/8 erroneous};
    \node[below] at (4,-0.22) {Heldout decision states (eight per condition)};
  \end{scope}
  \fill[fill=blue!10] (-3.7,4.45) rectangle (-3.45,4.65);
  \node[anchor=west,font=\footnotesize] at (-3.3600000000000003,4.55) {Authorized optimum};
  \fill[fill=red!10] (0.1,4.45) rectangle (0.35,4.65);
  \node[anchor=west,font=\footnotesize] at (0.44000000000000006,4.55) {Inadmissible};
  \fill[fill=gray!20] (3.2,4.45) rectangle (3.45,4.65);
  \node[anchor=west,font=\footnotesize] at (3.54,4.55) {Admissible, suboptimal};
\end{tikzpicture}

%% file: 3-assets/1-User/figures/nested-outcomes.tex
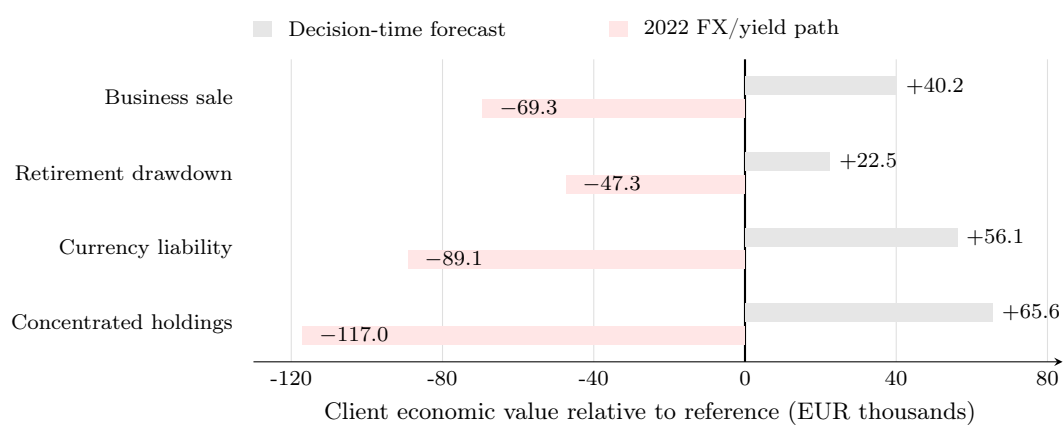
\begin{figure}[htbp]
\centering
\input{3-assets/1-User/figures/plots/nested-outcomes}
\caption{Forecast and subsequent client consequences on the same economic scale.}
\label{fig:nested-outcomes}
{\raggedright\fignote{One complete heldout case per family; all three instruction treatments select the same pair. Both bars measure payment-adjusted economic wealth plus constructed service value, relative to the reference, in EUR thousands. The forecast bar averages the three declared valuation scenarios. The 2022 bar uses historical FX and yield observations crossed with the same three constructed growth paths and weights (0.25, 0.50, 0.25). The plotted metric excludes the risk penalty used in decision-time client value. Source: frozen calculations and retained counterfactual cash flows.}}
\end{figure}

%% file: 3-assets/1-User/figures/plots/nested-outcomes.tex
\begin{tikzpicture}[font=\small,>=stealth]
  \draw[gray!25] (0.500,0.35) -- (0.500,4.35);
  \node[below,font=\footnotesize] at (0.500,0.35) {-120};
  \draw[gray!25] (2.500,0.35) -- (2.500,4.35);
  \node[below,font=\footnotesize] at (2.500,0.35) {-80};
  \draw[gray!25] (4.500,0.35) -- (4.500,4.35);
  \node[below,font=\footnotesize] at (4.500,0.35) {-40};
  \draw[gray!25] (6.500,0.35) -- (6.500,4.35);
  \node[below,font=\footnotesize] at (6.500,0.35) {0};
  \draw[gray!25] (8.500,0.35) -- (8.500,4.35);
  \node[below,font=\footnotesize] at (8.500,0.35) {40};
  \draw[gray!25] (10.500,0.35) -- (10.500,4.35);
  \node[below,font=\footnotesize] at (10.500,0.35) {80};
  \draw[thick] (6.5,0.35) -- (6.5,4.35);
  \draw[->] (0,0.35) -- (10.7,0.35);
  \node[left,align=right,font=\footnotesize] at (-.15,3.85) {Business sale};
  \fill[fill=gray!20] (6.50000,3.88) rectangle (8.50821,4.12);
  \node[right,font=\footnotesize] at (8.50821,4.00) {$+40.2$};
  \fill[fill=red!10] (3.03275,3.58) rectangle (6.50000,3.82);
  \node[anchor=west,text=black,font=\footnotesize] at (3.13275,3.70) {$ -69.3 $};
  \node[left,align=right,font=\footnotesize] at (-.15,2.85) {Retirement drawdown};
  \fill[fill=gray!20] (6.50000,2.88) rectangle (7.62358,3.12);
  \node[right,font=\footnotesize] at (7.62358,3.00) {$+22.5$};
  \fill[fill=red!10] (4.13346,2.58) rectangle (6.50000,2.82);
  \node[anchor=west,text=black,font=\footnotesize] at (4.23346,2.70) {$ -47.3 $};
  \node[left,align=right,font=\footnotesize] at (-.15,1.85) {Currency liability};
  \fill[fill=gray!20] (6.50000,1.88) rectangle (9.30515,2.12);
  \node[right,font=\footnotesize] at (9.30515,2.00) {$+56.1$};
  \fill[fill=red!10] (2.04415,1.58) rectangle (6.50000,1.82);
  \node[anchor=west,text=black,font=\footnotesize] at (2.14415,1.70) {$ -89.1 $};
  \node[left,align=right,font=\footnotesize] at (-.15,0.85) {Concentrated holdings};
  \fill[fill=gray!20] (6.50000,0.88) rectangle (9.77818,1.12);
  \node[right,font=\footnotesize] at (9.77818,1.00) {$+65.6$};
  \fill[fill=red!10] (0.64962,0.58) rectangle (6.50000,0.82);
  \node[anchor=west,text=black,font=\footnotesize] at (0.74962,0.70) {$ -117.0 $};
  \fill[fill=gray!20] (0,4.65) rectangle (.25,4.85);
  \node[anchor=west,font=\footnotesize] at (.34,4.75) {Decision-time forecast};
  \fill[fill=red!10] (4.7,4.65) rectangle (4.95,4.85);
  \node[anchor=west,font=\footnotesize] at (5.04,4.75) {2022 FX/yield path};
  \node[below] at (5.25,-.05) {Client economic value relative to reference (EUR thousands)};
\end{tikzpicture}

%% file: 3-assets/1-User/tables/narrative-live-results.tex
\begin{table}[htbp]
\centering
\caption{Initial narrative outcomes across three repetitions, with 42 records per workflow. Costs cover 37 matched records per workflow.}
\label{tab:narrative-live}
\small
\begin{tabular*}{\textwidth}{@{\extracolsep{\fill}}lrrr@{}}
\toprule
Measure & Deterministic & Single agent & Multi-agent \\
\midrule
Correct terminal decisions & 42/42 & 30/42 & 22/42 \\
Correct permitted recommendations & 24 & 14 & 7 \\
Correct required holds & 18 & 16 & 15 \\
Failed workflows & 0 & 12 & 20 \\
Inadmissible raw proposals / proposals & 0/54 & 1/37 & 0/20 \\
Reported model cost (USD) & 0.0000 & 7.7222 & 13.3133 \\
\bottomrule
\end{tabular*}
\end{table}

%% file: 3-assets/1-User/tables/narrative-interface-results.tex
\begin{table}[htbp]
\centering
\caption{Revised-interface outcomes across three repetitions, with complete accounting for all 42 records per workflow.}
\label{tab:narrative-interface}
\small
\begin{tabular*}{\textwidth}{@{\extracolsep{\fill}}lrrr@{}}
\toprule
Measure & Deterministic & Single agent & Multi-agent \\
\midrule
Technical completion & 42/42 & 42/42 & 42/42 \\
Correct terminal decisions & 42/42 & 42/42 & 42/42 \\
Correct permitted recommendations & 24 & 24 & 24 \\
Correct required holds & 18 & 18 & 18 \\
Failed workflows & 0 & 0 & 0 \\
Inadmissible raw proposals / proposals & 0/54 & 0/54 & 0/54 \\
Reported model cost (USD) & 0.0000 & 2.1314 & 5.6348 \\
\bottomrule
\end{tabular*}
\end{table}

%% file: 3-assets/1-User/figures/workflow-results.tex
\begin{figure}[htbp]
\centering
\input{3-assets/1-User/figures/plots/workflow-results}
\caption{Narrative conformance and revised model expenditure.}
\label{fig:workflow-results}
{\raggedright\fignote{Panel (a) shows observed conformance in the initial and revised narrative comparisons; panel (b) shows revised model expenditure with complete accounting. Sources: retained narrative comparisons summarised in \cref{tab:narrative-live,tab:narrative-interface}.}}
\end{figure}

%% file: 3-assets/1-User/figures/plots/workflow-results.tex
\input{3-assets/1-User/figures/plots/result-values.tex}
\begin{tikzpicture}[font=\small,>=stealth]
  \node[anchor=west,font=\small\bfseries] at (-2.35,4.65) {(a) Correct terminal decisions};
  \node[anchor=west,font=\small\bfseries] at (6.75,4.65) {(b) Revised model cost};
  \begin{scope}[x=0.112cm]
    \foreach \v in {0,14,28,42} {
      \draw[gray!25] (\v,0.5) -- (\v,4.1);
      \node[below,font=\footnotesize] at (\v,0.45) {\v};
    }
    \draw[->] (0,0.5) -- (46,0.5);
    \foreach \y/\name/\initial/\revised in {
      3.55/Deterministic/\InitialDeterministic/\RevisedDeterministic,
      2.35/Single agent/\InitialSingle/\RevisedSingle,
      1.15/Multi-agent/\InitialMulti/\RevisedMulti} {
      \node[left] at (0,\y) {\name};
      \fill[fill=gray!20] (0,\y+0.04) rectangle (\initial,\y+0.32);
      \fill[fill=blue!10] (0,\y-0.32) rectangle (\revised,\y-0.04);
      \node[right,font=\footnotesize] at (\initial,\y+0.18) {\initial};
      \node[right,font=\footnotesize] at (\revised,\y-0.18) {\revised};
    }
    \node[below] at (21,-0.02) {Decisions matching the reference (of 42)};
  \end{scope}
  \fill[fill=gray!20] (0,4.12) rectangle (0.28,4.32);
  \node[anchor=west,font=\footnotesize] at (0.34,4.22) {Initial};
  \fill[fill=blue!10] (1.9,4.12) rectangle (2.18,4.32);
  \node[anchor=west,font=\footnotesize] at (2.24,4.22) {Revised};
  \begin{scope}[shift={(8.8,0)},x=0.72cm]
    \foreach \v in {0,2,4,6} {
      \draw[gray!25] (\v,0.5) -- (\v,4.1);
      \node[below,font=\footnotesize] at (\v,0.45) {\v};
    }
    \draw[->] (0,0.5) -- (6.6,0.5);
    \foreach \y/\name/\cost/\label in {
      3.55/Deterministic/\CostDeterministic/\CostLabelDeterministic,
      2.35/Single agent/\CostSingle/\CostLabelSingle,
      1.15/Multi-agent/\CostMulti/\CostLabelMulti} {
      \node[left] at (0,\y) {\name};
      \fill[fill=gray!20] (0,\y-0.2) rectangle (\cost,\y+0.2);
      \node[right,font=\footnotesize] at (\cost,\y) {\label};
    }
    \node[below] at (3,-0.02) {USD per 42 assignments};
    \node[anchor=west,font=\footnotesize,text=red!60!black] at (0,4.22)
      {Multi/single: $\CostRatio\times$};
  \end{scope}
  \node[anchor=north west,align=left,font=\footnotesize,text width=15.9cm] at (-2.35,-0.72)
    {14 authored cases, three repetitions; revised interface evaluated on the same cases.\\
     Every revised calculation had one feasible candidate. Cost covers provider model charges.};
\end{tikzpicture}